\documentclass[preprint]{aastex}

\def\kmps{\ifmmode {\rm\,km\,s^{-1}}\else 
             ${\rm\,km\,s^{-1}}$\fi}
\newcommand{\kms}{\kmps}
\def\ergcms{\ifmmode {\rm\,ergs\,cm^{-2}\,s^{-1}}\else
             ${\rm\,ergs\,cm^{-2}\,s^{-1}}$\fi}
\newcommand{\fluxunit}{\ergcms}
\def\ergcmsa{\ifmmode {\rm\,ergs\,cm^{-2}\,s^{-1}\,\AA^{-1}}\else
             ${\rm\,ergs\,cm^{-2}\,s^{-1}\,\AA^{-1}}$\fi}

\def\ergs{\ifmmode {\rm\,ergs\,s^{-1}}\else
             ${\rm\,ergs\,s^{-1}}$\fi}
\newcommand{\lum}{\ergs}
\def\lya{\ifmmode {\mbox{Ly}\alpha}\else
             {Ly$\alpha$}\fi}
\def\mdot{\ifmmode {\, M_\odot\, \rm yr^{-1}}\else
             $\, M_{odot}\,\rm yr^{-1}$\fi}

\slugcomment{ver. 7 May 2007}
\shorttitle{Spectroscopy of Extended \lya\ Sources}
\shortauthors{Saito et al.}

\begin{document}

\title{
Deep Spectroscopy of Systematically Surveyed 
Extended Lyman-$\alpha$ Sources at $z\sim 3-5$
\footnote{Based on observations made with ESO telescope at 
the Paranal Observatory, under the programme ID 074.A-0524}
\footnote{Based on data collected at the Subaru Telescope, 
which is operated by the National Astronomical Observatory of Japan.}
}

\author{
Tomoki Saito\altaffilmark{3,4}, Kazuhiro Shimasaku\altaffilmark{3}, 
Sadanori Okamura\altaffilmark{3}, Masami Ouchi\altaffilmark{5,6}, 
Masayuki Akiyama\altaffilmark{7}, 
Michitoshi Yoshida\altaffilmark{8}, 
Yoshihiro Ueda\altaffilmark{9}
}
\altaffiltext{3}{Department of Astronomy, University of Tokyo, 
    7-3-1 Hongo, Bunkyo, Tokyo 113-0033, Japan}
\altaffiltext{4}{Present Address: 
Dark Cosmology Centre, Niels Bohr Institute, University of Copenhagen, 
Juliane Maries Vej 30, 2100 Copenhagen \O, Denmark, 
e-mail: tomoki@dark-cosmology.dk}
\altaffiltext{5}{Space Telescope Science Institute, 
3700 San Martin Drive, Baltimore, MD 21210, USA}
\altaffiltext{6}{Hubble Fellow}
\altaffiltext{7}{Subaru Telescope, National Astronomical Observatory of Japan, 
North A'ohoku Place, Hilo, HI 96720, USA}
\altaffiltext{8}{Okayama Astrophysical Observatory, 
National Astronomical Observatory of Japan}
\altaffiltext{9}{Department of Astronomy, Kyoto University , 
Sakyo-ku, Kyoto 606-8502, Japan}

\begin{abstract}
Spatially extended \lya\ sources that are faint and/or compact 
in continuum are candidates for extremely young ($\lesssim 10^7\rm yrs$) 
galaxies at high redshifts. 
We present medium-resolution ($R\sim 2000$) spectroscopy of 
such extended \lya\ sources found in our previous study 
at $z\sim 3-5$, using VLT/VIMOS. 
The deep spectroscopy showed that all 18 objects we observed 
have large equivalent widths (EWs) exceeding 100\AA. 
For about 30\% of our sample (five objects), 
we identified conspicuous asymmetry on the profiles of the \lya\ line. 
They show broad wing emission components on the red side, 
and sharp cut-off on the blue side of the \lya\ line. 
Such asymmetry is often seen in superwind galaxies known to date, 
and also consistent with a theoretical prediction of superwinds. 
In fact, one of them show systematic velocity structure on the 
2-dimensional spectrum suggesting the existence of superwind activity. 
There are eight objects ($8/18\sim 40\%$) that have large EWs exceeding 
200\AA, and no clear signature of superwind activities. 
Such large EWs cannot be explained in terms of photo-ionization 
by a moderately old ($>10^7\rm yrs$) stellar population, 
even with a top-heavy IMF or an extremely low metallicity. 
These eight objects clearly show a positive correlation between 
the \lya\ line luminosity and the velocity width. 
This suggests that these eight objects are good candidates for 
forming-galaxies in a gas-cooling phase. 
\end{abstract}



\section{Introduction}
\label{introduction}
Galaxies undergoing their initial star formation are predicted 
to have strong \lya\ emission \citep{partridge, cf93}. 
Recent imaging surveys using narrowband filters tuned to the 
redshifted \lya\ line have successfully identified a large 
number of \lya\ emitters 
\citep[LAEs, e.g. ][]{ch98, hu98, fynbo, fynbo03, rhoads00, rhoads03, 
ouchi03, ouchi05, shimasaku03, kodaira, maier, dawson04, kashikawa, iye06}. 
Some of them are fairly faint in rest-frame UV continuum and have 
large \lya\ equivalent widths exceeding 240\AA\ \citep[hereafter MR02]{mr2002}. 
Such large equivalent widths suggest that they are possibly 
in the initial phases ($\ll 10^7 \rm yrs$) of star formation. 
The faintness in UV continuum also suggests that their stellar 
masses are fairly small, implying that a significant star formation 
has not yet occurred. 

In such early phases of galaxy formation (age of $\lesssim 10^7\rm yrs$), 
(proto)galaxies are predicted to radiate spatially extended \lya\ 
emissions. At least three mechanisms are currently 
proposed for extended \lya\ emissions; {\it i}) cooling radiation 
from gravitationally heated primordial gas infalling into a dark halo 
potential 
\citep{haiman00, fardal, furlanetto04, bd03, db06, dijkstra06a, dijkstra06b}, 
{\it ii}) resonant scattering of \lya\ photons from the central 
(hidden) ionizing source 
\citep{moller98, hr01, weidinger04, weidinger05, laursen06}, 
{\it iii}) starburst-driven galactic winds 
\citep{ts00, tsk01, ohyama, geach05}. 

Recent narrowband imaging surveys for high-$z$ \lya\ emitters (LAEs) have 
found several tens of extended \lya\ sources with no 
significant UV continuum sources sufficient to ionize such a large 
amount of \ion{H}{1} gas 
\citep[\lya\ blobs, or LABs:][]{keel99,s00,francis01,m04,palunas04,dey05,nilsson}
However, there are still only restricted number of extended \lya\ sources 
known to date, while thousands of normal LAEs are already known. 
Most of the extended \lya\ sources known to date are identified in 
narrowband imaging surveys, which cover very narrow redshift range. 
Even the largest sample made by Matsuda et al. (hereafter M04), 
which is constructed from a narrowband survey targeted on LABs 
of \citet[hereafter S00]{s00}, 
covers a redshift range of only $z=3.06^{+0.04}_{-0.03}$. 

Due to the lack of a systematic sample and subsequent follow-up studies, 
nature of such extended \lya\ sources remains unknown. 
Even for the most famous objects, two LABs of S00, 
the origins of the extended \lya\ emission are not fully clear. 
For example, observational studies 
of S00's LABs are made by many astronomers, ranging from radio to 
X-ray wavelengths \citep[e.g.][]{chapman01,chapman04,bzs04,geach05,m06}. 
These studies suggest that obscured ionizing sources such as starburst 
regions or AGNs are associated with LABs. 
Detailed optical spectroscopy of the LABs suggests that 
outflow from the central sources are responsible for the extended 
\lya\ emission \citep{ohyama, wilman}, while numerical simulation 
shows that cooling radiation from infalling gas combined with 
a central ionizing source can also explain the spatial extent, 
line profiles, and velocity widths of the \lya\ emission 
\citep{dijkstra06b}. There is also a detailed observational study 
by \citet{bower04} suggesting that extended \lya\ emission of LAB 
cannot be simply explained with a single mechanism such as a 
galactic wind or cooling radiation. 
They are likely to be massive systems harboring galaxy formation 
sites \citep{m06}, although the true nature of the \lya\ emission 
are still unclear. 

In order to construct a systematic sample of extended \lya\ sources covering 
a wide range of redshift, we have carried out a deep, wide-field 
imaging survey of a blank-field, Subaru/{\it XMM-Newton} Deep Field, 
using Subaru Telescope and seven intermediate-band (IA) filters. 
This survey enabled us to construct a sample of 41 extended \lya\ sources 
located at $z\sim 3-5$, and also to show that this kind of objects 
commonly exist in the early universe far beyond $z\sim 3$ 
\citep[][hereafter Paper1]{paper1}. 
We here present a deep follow-up spectroscopy of our photometric 
sample of Paper1 using high-resolution grism of VIMOS on VLT. 
This gives a three times deeper exposure and a several times higher 
resolution than our previous spectroscopy presented in Paper1. 
We constructed a spectroscopic sample of 18 spatially extended \lya\ 
sources with large equivalent widths, located at redshifts 
of $3.3\lesssim z\lesssim 4.7$. This sample allows us to make a 
quantitative analysis in terms of equivalent width, velocity width, 
and \lya\ line luminosity, as well as the velocity structure traced 
by the \lya\ line profile. 

The rest of this paper is organized as follows. 
We briefly introduce our photometrically selected sample of 
extended \lya\ sources in \S\ref{sample}, based on the previous 
observations in Paper1. 
In \S\ref{spec} we present the details of our spectroscopic observations 
and data reduction. 
The results of the spectroscopy thus obtained and a discussion based on 
them are presented in \S\ref{res}. 
Finally, we summarize the conclusions in \S\ref{con}. 

All magnitudes are in the AB \citep{oke, fsi95}. 
We use the standard $\Lambda$-dominated CDM cosmology with 
$\Omega_M = 0.3$, $\Omega_\Lambda = 0.7$, 
and $H_0 = 100h = 70\kms\rm Mpc^{-1}$, unless otherwise noted.

\section{Our sample of extended \lya\ sources}
\label{sample}
The targets of our current work were selected from the photometric 
sample constructed in Paper1, based on an intermediate-band imaging survey 
(Subaru proposal ID S02B-163: Kodaira et al.) with Subaru Prime-focus 
Camera (Suprime-Cam) mounted on 8.2m Subaru Telescope. 
The field of this survey is a blank field called Subaru/{\it XMM-Newton} 
Deep Field South (SXDF-S) located at $\alpha = 02^h18^m00^s$, 
$\delta = -05^\circ 25'00''$ (J2000). 
The selection of extended \lya\ sources was based on photometry 
in seven intermediate (IA) bands centered at $\sim 5000-7100\rm\AA$ 
\citep{hayashino}, the $B$ band, and a broadband redward of the 
\lya\ line ($R$, $i'$, or $z'$ band). 
The broadband data were taken in a large project of Subaru Telescope, 
Subaru/{\it XMM-Newton} Deep Survey \citep[SXDS: ][]{sekiguchi04}. 
The details of the selection are reported in Paper1. 
This sample consists of 41 objects, all of which are basically 
\lya\ emitters with $B$-dropout, located at redshifts of 
$3.24\lesssim z\lesssim 4.95$. 
Their photometric properties are listed in Table 3 of Paper1. 

Their \lya\ line luminosities measured with the IA-band photometry 
are typically $\sim 10^{42}-10^{43}\lum$. 
They are spatially extended in \lya\ line, and faint and/or compact 
in the continuum redward of the \lya. 
They can be divided into two minor classes; 
``continuum-compact'' objects and ``continuum-faint'' objects. 
About a half of them, 22 objects, can be regarded as point-sources 
in the redward continuum, and are classified as continuum-compact objects. 
They show $B$ band dropout, which is consistent with Lyman break feature 
of distant galaxies at $z\gtrsim 3$. The remaining 19 objects are 
continuum-faint objects, which are fainter than $3\sigma$ in the 
redward continuum band. 
Their continuum break between the redward band and the $B$ band 
cannot be measured since they are not detected in the $B$ band. 
Some of these continuum-faint objects appear to be larger than 
the PSF size in the redward continuum band, 
but the sizes measured here are not reliable due to the low signal-to-noise 
(S/N) ratio. 

\begin{figure}
\plotone{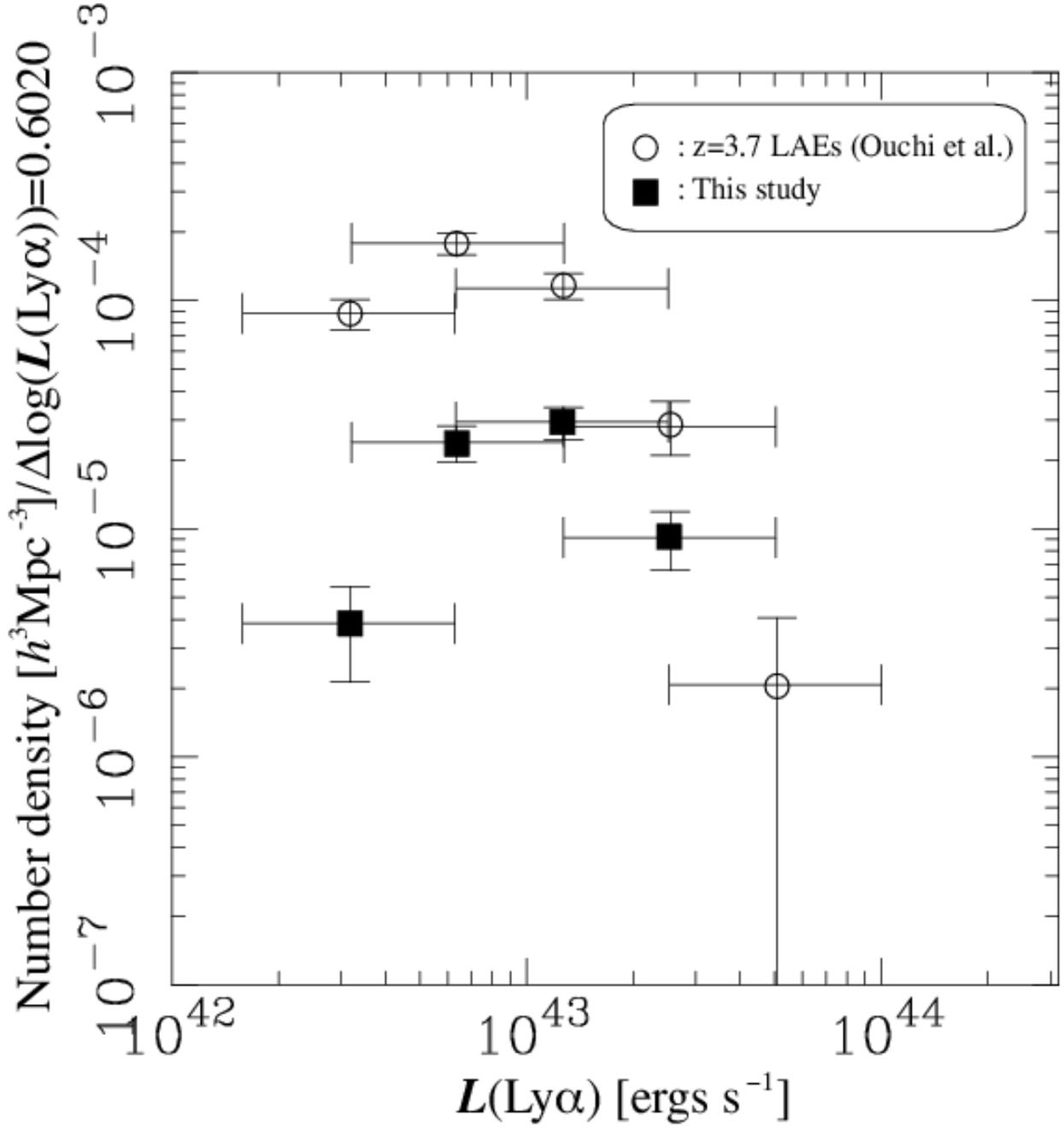}
\caption{Luminosity functions (LFs) of our sample (filled squares) 
and a sample of normal LAEs at $z\simeq 3.7$ by Ouchi et al. (open circles). 
The vertical error bars show the Poisson errors, and the horizontal 
error bars represent the size of the luminosity bins. 
}
\label{fig:lf}
\end{figure}

Since our IA bands have relatively low sensitivities to the line emission 
compared with usual narrowbands, we can select objects with large equivalent 
widths (EWs). Our selection criteria correspond to the EWs 
greater than $\sim 55\rm\AA$ in the rest frame. 
We roughly estimated the EWs with the photometric data, and found 
that the largest value is $\sim 700\rm\AA$. 
The \lya\ line luminosities are also relatively large, 
ranging from several $\times 10^{42}\lum$ to $\sim 1\times 10^{43}\lum$. 
This corresponds to the luminous-end of the luminosity function (LF) 
of normal LAEs at similar redshifts. 
Figure \ref{fig:lf} shows a comparison of the LF between our sample 
and a sample of LAEs at $z\simeq 3.7$ from a narrowband survey 
of SXDF (Ouchi et al. in preparation). 
The completeness of Ouchi et al.'s is $\sim 70-80\%$ at $\le 24.2$ mag 
($\gtrsim 3.4\times 10^{42}\lum$) and $\sim 50\%$ at $\le 24.7$ mag 
($\gtrsim 5.4\times 10^{42}\lum$). 
This comparison may suggest that a significant fraction ($\sim 10-20\%$) 
of luminous LAEs are spatially extended \lya\ sources. 
However, one object in our sample at $z\simeq 3.7$, 
IB10-90651, is not included in this LAE sample down to 24.7 mag. 
This is due to the low surface brightness of this object and the 
relatively shallow exposure of the narrowband image. 

\begin{figure}
\plotone{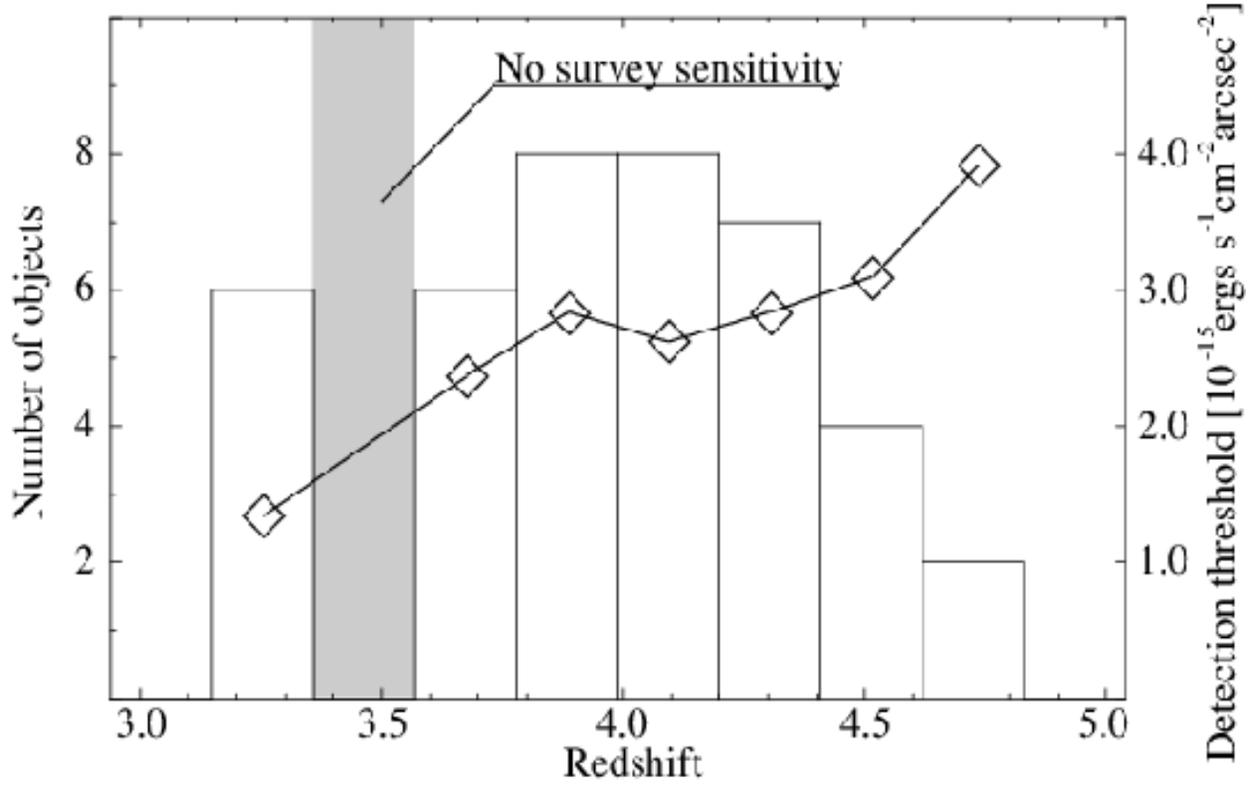}
\caption{Redshift distribution of extended \lya\ sources. 
The histogram shows the redshift distribution of our photometric 
sample of 41 extended \lya\ sources, based on coarse redshift 
information obtained from the IA imaging (redshift resolution 
$\Delta z\sim 0.2$). The open diamonds show the detection limit 
($2\sigma$) for each IA band in units of rest-frame surface brightness 
(right axis). 
}
\label{fig:z-hist}
\end{figure}

\begin{figure}
\plotone{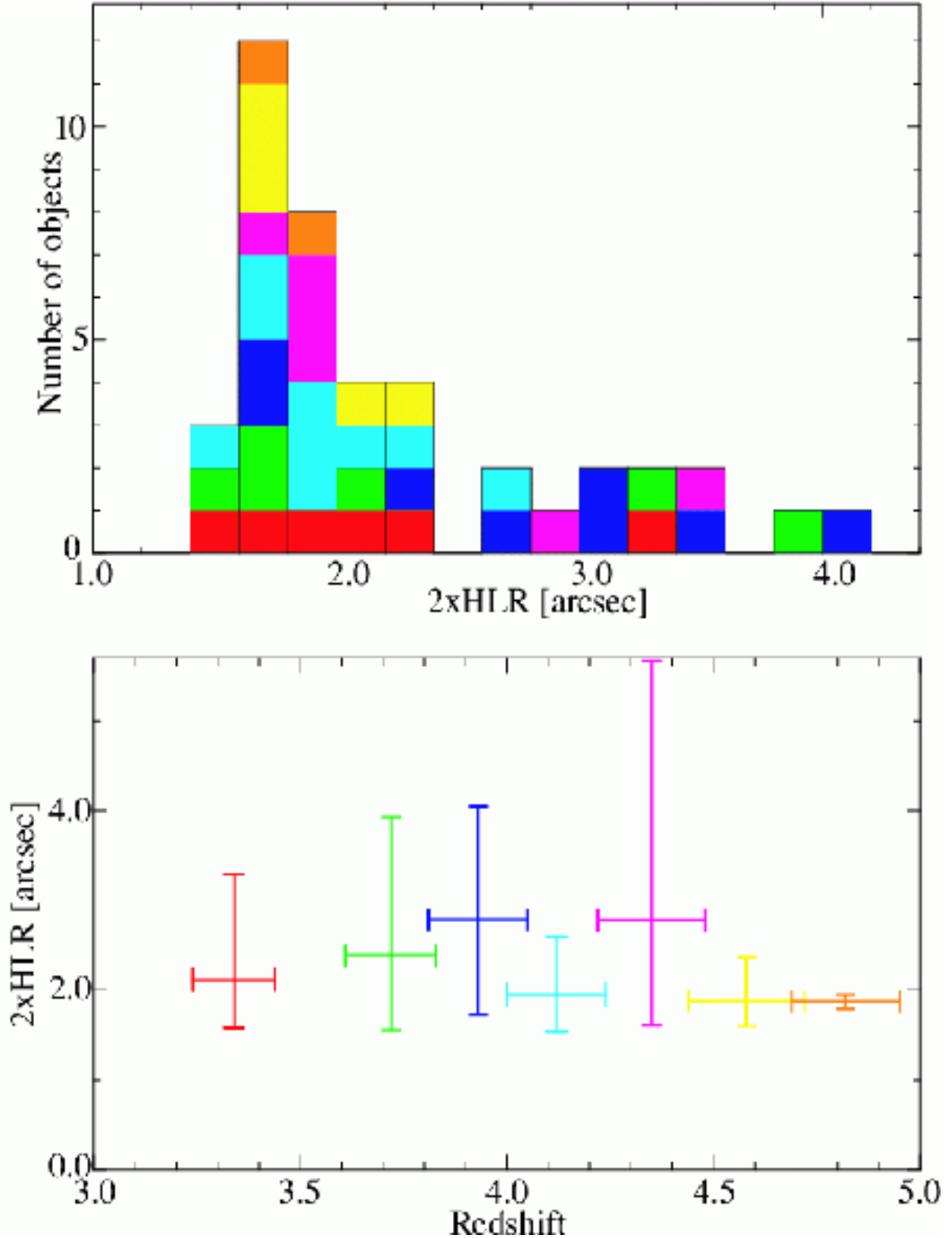}
\caption{Size distribution of the photometric sample of 41 extended 
\lya\ sources. The top panel is a histogram of the size defined by 
twice the half-light-radius (HLR). The bottom panel shows the size 
as a function of (roughly estimated) redshift. 
For both panels, colors correspond to the bands 
where we detected the \lya\ emission. In the bottom panel, 
the horizontal error bars indicate the bandwidths of the IA bands, 
and the vertical error bar shows the minimum and maximum value. }
\label{fig:sizedist}
\end{figure}

Figure \ref{fig:z-hist} shows the redshift distribution of our objects, 
based on the IA imaging (two double-counted objects are not eliminated). 
The detection limit of each IA band is also plotted in this figure. 
This shows that extended \lya\ sources commonly exist in the early 
universe, almost independently of redshift. 
The size distribution of our objects is shown in Figure \ref{fig:sizedist}. 
There is no significant redshift dependence in the spatial extent. 
Note that the low sensitivity to the line emission also affects 
the apparent spatial extents of our objects, especially in higher 
redshift bands. The diffuse emission at the outskirts becomes more 
difficult to be detected, and the apparent sizes tend to become smaller. 
This is also suggested from Figure \ref{fig:z-hist}, in which the number of 
objects declines with redshift. 
True sizes may be larger than our estimation, and the size distribution 
may resemble that of M04's LAB sample. 
Detailed simulations of such effects are presented in Paper1. 

\begin{figure}
\plotone{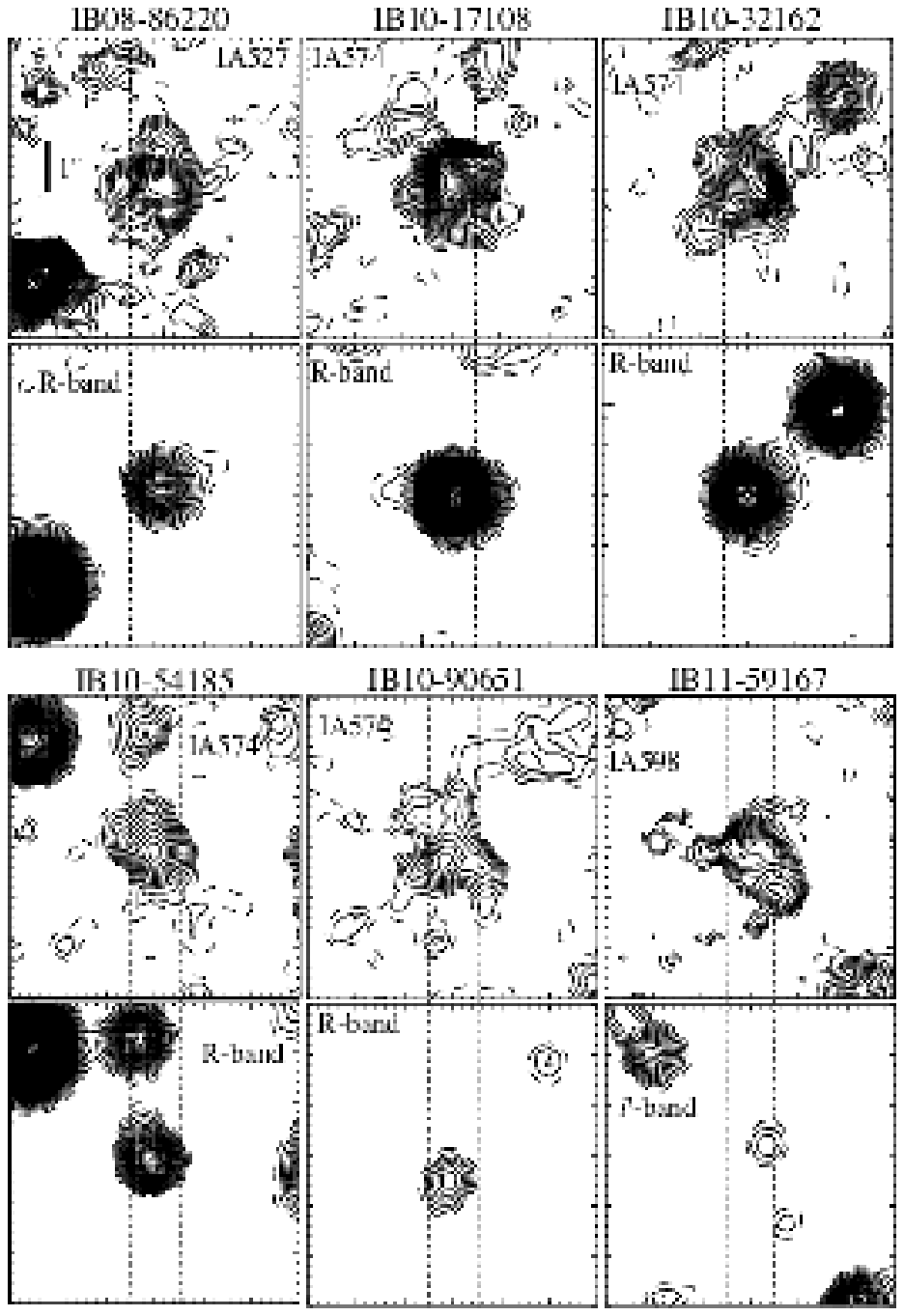}
\caption{Images of the 18 extended \lya\ sources. 
For each object, the top panel shows the IA band image where the \lya\ line 
is detected, and the bottom panel shows the redward broadband image. 
The vertical dotted lines show the slit positions. 
}
\label{fig:images}
\end{figure}

From the sample of 41 photometrically-selected objects, 
we selected 21 objects for further follow-up spectroscopy described below. 
About a half of them (11 objects) are continuum-compact objects, 
and another half (10 objects) are continuum-faint obejcts. 
As noted below, we could not obtain the spectra of three objects: 
IB11-101786 (continuum-faint), IB13-62009 (continuum-faint), and 
IB14-62116 (continuum-compact). Excluding these three, 
we obtained a spectroscopic sample of 18 objects (10 continuum-compact, 
and 8 continuum-faint objects). 
Photometric properties of the 18 objects are briefly summarized in 
Table \ref{tab:phot}, and their images are shown in Figure \ref{fig:images}. 
Seven of them were spectroscopically confirmed to be high-$z$ \lya\ 
emitters (Paper1).

\section{VIMOS spectroscopy}
\label{spec}
\subsection{Observations}
\label{spec:obs}
We used Visible Multi Object Spectrograph (VIMOS) mounted on the 8.2m 
Very Large Telescope (VLT) UT3 ``Melipal'' to take deep spectra 
for the objects selected above, under the program ID 074.A-0524. 
The observations were made during three dark nights of the visitor 
mode run on 2004 November 6 to 9 (UT). 
Two masks were used in multi-object spectroscopy (MOS) mode of VIMOS, 
covering the 21 objects. Two of them, IB11-101786 and IB13-62009, were 
too faint to be detected with the current depth of our spectroscopy. 
For another one object, IB14-62116, the \lya\ line was almost completely 
overlapped with a strong sky emission line that cannot be correctly 
subtracted. We thus obtained spectra for 18 objects. 
In this run, the HR-Orange grism and the GG435 order-cut filter were used. 
This setting gives a spectral resolution 
$R=\Delta\lambda/\lambda \approx 2160$ ($\Delta\lambda \approx 2.8\rm\AA$ 
at $\lambda = 6000\rm\AA$) with a $1''.0$ wide slit. 
This spectral resolution corresponds to a velocity resolution 
of $\approx 140\kms$. 

VIMOS has a wide FoV consisting of four quadrants of $\sim 7'\times 8'$ 
each, which can cover about 1/3 of a single Suprime-Cam FoV. 
The sky distribution of our objects and the VIMOS FoV are shown 
in Figure \ref{fig:skydist}. 
The two MOS masks cover ten and eight objects, respectively. 
The integration time was 9.5 hrs and 8.0 hrs, respectively. 
During this run, the shutter unit for the second (north-eastern) 
quadrant often did not work correctly, and several exposures 
(30 minutes each) were lost. As a consequence, the integration 
time for each quadrant ranges from 6.5 hrs to 9.5 hrs. 
The seeing size varied during the three nights from $\lesssim 0''.5$ to 
$\sim 2''.4$. 
We flagged the frames with seeing worse than $\sim 1''.0$, so that 
the effective integration times of objects 
range from 4.0 hrs to 7.0 hrs. 
e standard star frames were taken at the end of the run. 
We used LTT3864, an F-dwarf located at $\alpha = 10^h 32^m 13.90^s$, 
$\delta = -35^\circ 37' 42''.4$ (J2000) as a standard star. 

\subsection{Primary reduction and flux measurement}
\label{spec:flux}
The primary reduction was made with the VIMOS pipeline recipes 1.0 
provided by ESO. This software package performs flat-fielding, 
distortion correction, wavelength calibration, and sky subtraction. 
The spectral resolution was measured to be $R\approx 2200$, 
using the sky emission lines. This resolution is sufficient to resolve 
the \lya\ line profiles of our objects, which are known to have velocity 
widths less than several $100\kms$ (\S 4.2 of Paper1). 
After primary reduction using the VIMOS pipeline, we performed 
flux calibration and further analysis using our original software. 
For the two-dimensional spectra, we smoothed the data with Gaussian 
kernel with FWHM of $4\times 5$ pixels, which corresponds, to 
$3.22\rm\AA\, (\mbox{wavelength axis}) \times 
1''.0\, (\mbox{spatial axis})$. 
To measure the fluxes and the velocity width on the one-dimensional 
spectra, we used 5-pixel smoothing, roughly corresponding to the 
current spectral resolution. 
We then measured the central wavelengths, fluxes, and velocity 
widths of the \lya\ lines. 
The spectra thus obtained are shown in Figure \ref{fig:spec}. 

\subsection{Luminosity and velocity width of the \lya\ line}
\label{spec:lum-dv}
The \lya\ fluxes were calculated by correcting for the fraction 
of their line fluxes collectable with the $1''.0$ wide slitlets 
(slit-loss correction). 
The slit-loss correction was made by the following procedure by 
combination of photometric and spectroscopic data. 
We first assumed that (1) the intrinsic continuum spectra are flat within 
the bandpass of the IA filters, and that 
(2) the continuum levels are equal to that measured with the 
broadband redward of the \lya\ line. 
Then we calculated the effect of the IGM absorption to the continuum 
using the redshift information obtained with the spectroscopy 
by following the formulation of \citet{madau95}. 
The pure \lya\ line fluxes can be estimated by subtracting the continuum 
contribution from the IA photometry. 
We made this estimation of the total flux within an automatic aperture, 
$F_{total}$, and the flux within the slit we used, $F_{slit}$. 
The fraction of the collectable flux was then estimated by taking the ratio 
of the flux values measured with the both apertures, i.e., the slit-loss 
correction was applied by multiplying the line fluxes (from spectra) 
with $\frac{F_{total}}{F_{slit}}$. 

The \lya\ line fluxes thus obtained, $F\rm(Ly \alpha)$, are shown in 
Table \ref{tab:spec}, 
together with the values before the slit-loss correction. 
The errors listed in Table \ref{tab:spec} are estimated by integrating 
the $1\sigma$ noise level of the spectra over the range we integrated 
the flux of the \lya\ line. 

The \lya\ line luminosities were directly 
calculated from these flux values by using the redshifts obtained 
spectroscopically. We fitted a Gaussian function to the line profile 
of each object, and defined the line center as the central wavelength 
of the Gaussian. Then we determined the redshift from the wavelength of 
the line center, and calculated the luminosity, $L\rm(Ly\alpha)$, from 
the flux and the luminosity distance at the redshift. 
The velocity widths (FWHM) were measured simultaneously by using the 
redshifts obtained by the Gaussian fitting. 
We defined the zero velocity as the line center described above, 
and calculated the line-of-sight velocities of the wavelengths 
at which the flux density becomes a half of the peak value. 
The upper and lower limits of the FWHM were estimated by measuring the 
full velocity widths at $\mbox{(half maximum)}\pm 1\sigma$. 

\subsection{Equivalent widths of the \lya\ line}
\label{spec:ew}
Since most of our objects do not show any continuum emission, 
we cannot obtain their EWs solely from the spectral data. 
Instead, we used the photometric data (broadband redward of the \lya\ line) 
to estimate the continuum flux densities. 
The continuum flux densities were estimated using a $2''\phi$ aperture. 
Since we measure the EWs using the total flux of the \lya\ line, 
this may cause an underestimate of the UV flux densities. 
For continuum-compact objects, this does not affect the results 
since the spatial extent is small enough. For continuum-faint objects, 
the spatial extents are generally larger than the PSF size (see Fig.2 
of Paper1). However, they are fairly faint in continuum ($<3\sigma$) and 
the measurement of their size is not so reliable. 
This implies that using a larger aperture make the measurement 
more sensitive to the sky fluctuation. 

Then we divided the \lya\ line fluxes obtained spectroscopically 
(slit-loss corrected) with the continuum 
fluxes obtained photometrically. Their rest-frame EWs, $\rm EW_{rest}$, 
are also shown in Table \ref{tab:spec}. 
All the objects in our sample have fairly large EWs with a median 
value of $\rm EW_{rest}\approx 210\AA$. 
For 9 out of 18 objects, the $\rm EW_{rest}$ exceed 200\AA. 
Eight out of these nine objects are continuum-faint objects, 
having yet larger EWs exceeding 240\AA, 
i.e., all of the continuum-faint objects in our VIMOS sample have large EWs. 
Another object with $\rm EW_{rest}>200\AA$, IB12-58572, 
is a continuum-compact object, but its continuum level is the 
lowest among all the continuum-compact objects in our VIMOS 
sample ($\sim 3.5\sigma$). We plot the EWs as a function of 
the \lya\ line luminosity in Figure \ref{fig:lew}. 
The nine objects with $\rm EW_{rest}>200\AA$ apparently show 
a positive correlation between the luminosity and the EW. 
This is thought to be an artifact caused by the detection limits 
in the redward continuum bands. For these nine objects (or at least 
eight continuum-faint objects), the EWs we measured are likely 
to be still larger, since the continuum levels are close to the 
sky fluctuation level. For the remaining nine objects, our measurement 
of the EWs is somewhat more accurate than for objects with large EWs, 
since the continuum levels exceed $3.7\sigma$. 

The upper and lower limits for the EWs are estimated as follows. 
Since the denominators in the calculation of EWs are very 
small values, and the errors in EWs are dominated by the photometric 
errors in the broadband images. 
We estimated the error using the limiting magnitudes of the broadband 
images used to estimate the continuum levels: 
27.4 mag for $R$ and 27.0 for $i'$ ($3\sigma$, $2''\phi$ aperture). 
The upper and lower limits of the EWs, $\rm EW_{upper}$ and $\rm EW_{lower}$, 
were estimated from 
\begin{displaymath}
{\rm EW_{upper}} = (F(\lya)+{\rm err}[F(\lya)])
/(f_\lambda - {\rm err}[f_\lambda]), 
\end{displaymath}
and 
\begin{displaymath}
{\rm EW_{lower}} = (F(\lya)-{\rm err}[F(\lya)])
/(f_\lambda + {\rm err}[f_\lambda]), 
\end{displaymath}
where $F(\lya)$ is the \lya\ line flux, $f_\lambda$ is the continuum 
flux density, and err[] denotes the 1$\sigma$ noise level. 

Note that our measurements of EWs are likely to underestimate the 
intrinsic value, since we did not make any correction for the 
IGM absorption of the \lya\ emission. 

\subsection{Wing component analysis}
\label{spec:wing}
Although most of our objects have relatively small velocity widths, 
some objects have broad wing emission of the \lya\ line. 
The \lya\ line profile of IB10-90651, for example, has broad wing 
emission on the red side, while the blue side shows a relatively 
sharp cut-off. Such features of the line profile must reflect the 
kinematics of \lya\ emitting gas, and should act as a diagnostic 
of inflow / outflow activities in the system. 
In order to identify such high velocity wing components in the \lya\ 
line profiles, we performed a quantitative analysis for all the objects 
in our spectroscopic sample. 
To identify such faint wing components, we need to have spectra of 
sufficiently high S/N ratios. 
We therefore used 8-pixel smoothing after integrating the spectra 
along the slit direction. Although the 8-pixel smoothing reduces the 
spectral resolution to $\sim 5\rm\AA$, this resolution is still high 
enough to investigate the gas dynamics with the line profile. 

The procedure of the analysis is similar to that described in 
\S\ref{spec:lum-dv}, 
except the smoothing width of 8-pixel. After obtaining improved S/N ratios 
by smoothing the one-dimensional spectra, we fitted a simple Gaussian 
function to the \lya\ line profile of each object. We then plotted the 
excess of the spectrum from the Gaussian function against the line-of-sight 
velocity. 
If the excess of the spectrum greater than $2\sigma$ has 
a velocity width greater than $250\kms$ (corresponds to 8-pixels), 
we here refer to it as the ``excess''. If the excess appears from 
within the $2\sigma$-width of the Gaussian peak and continuous 
beyond $\pm 500\kms$ from the line center, we define it as ``wing emission''. 
The plots thus obtained are shown in Figure \ref{fig:spec}, 
and the results are also summarized in Table \ref{tab:spec}. 
For at least five objects, IB10-17108, IB10-32162, IB10-54185, IB10-90651, 
and IB14-52102, we identified high velocity wing emission on the 
red side of the \lya\ line. On the other hand, none of our objects 
shows significant wing emission on the blue side. 
Just two objects, IB12-30834 and IB12-58572, {\em marginally} show 
a blue wing. 

\section{Results and discussion}
\label{res}
Summarizing the results obtained in the previous section, 
we categorize our objects according to the spectral properties. 
We can classify our objects in two ways, i.e., in terms of EWs 
and wing emission in the line profile. 
In terms of EWs, we divide our VIMOS sample into two EW classes. 
The moderate EW class has $\rm 100\AA < EW <200\AA$ while the 
large EW class has $\rm EW>200\AA$ (see Figure \ref{fig:lew}). 
The value of $\rm EW=200\AA$ is fairly large for ordinary starbursts. 
The large EW class and the small EW class roughly correspond to 
continuum-compact objects and continuum-faint objects, respectively. 
In terms of the line profile, we can also classify our objects into 
two groups, i.e., those with broad asymmetric wing on the red side, 
and those with no clear signature of such wing emission. 

\subsection{Objects with moderate EWs}
\label{res:SB}
The nine objects in the smaller EW group, IB08-86220, IB10-17108, 
IB10-32162, IB10-54185, IB12-21989, IB12-71781, IB13-96047, 
IB14-47257, and IB14-52102, have EWs ranging between 100\AA\ and 
200\AA. This EW range is the regime of stellar photo-ionization 
with a solar metallicity, an upper mass cut-off of $100M_\odot$, 
and an IMF slope up to 1.5 \citep{cf93}. 
The EWs of these objects thus can be explained in terms of 
photo-ionization by moderately old stellar populations, 
not necessarily requiring an extremely top-heavy IMF or a zero-metallicity. 
The most conservative explanation for these objects are thus 
ordinary starbursts. The large spatial extent of \lya\ emission 
can be understood as a result of resonant-scattering of \lya\ 
photons \citep[e.g.][]{moller98, laursen06}. 

Assuming that the ionizing sources for these objects are 
ordinary starburst. we can give rough estimates of the star 
formation rates (SFRs) of these objects. 
We here assume Salpeter's IMF, stellar mass range of $0.1-100M_\odot$, 
solar metallicity, no extinction, and the case B recombination 
in the low-density limit ($N_e\ll 1.5\times 10^4\rm cm^{-3}$). 
We can then use a conversion law of 
$L(\lya)=1\times 10^{42}({\rm SFR}/M_\odot\rm yr^{-1}\lum$) 
to estimate their SFRs \citep{agn2,kennicutt98}. 
Their SFRs thus estimated are $3.3-13\mdot$, which are located 
within the range of normal LAEs known to date 
\citep[e.g.][]{fynbo03,rhoads03,dawson04}. 

However, their physical nature, even the SFR, is poorly constrained 
because of the large uncertainties in dust / IGM absorption, 
stellar mass function, and the contribution from non-stellar processes. 
Some of the objects indeed show asymmetric line profiles suggesting 
superwind activities (see \S\ref{res:wing} below). 
The possibility of the existence of AGNs, although unlikely, 
also cannot be eliminated completely. 
Notes for individual objects in this group are summarized 
in appendix \ref{notes:modEW}.

\subsection{Objects with broad asymmetric wings}
\label{res:wing} 
As described in \S\ref{spec:wing}, five objects, 
IB10-17108, IB10-32162, IB10-54185, IB10-90651, and IB14-52102, 
have asymmetric broad wing emission on the \lya\ line 
profile. 
These objects have conspicuous high velocity wing components 
on the red side, and a relatively sharp cut-off on the blue side 
of the \lya\ line. 
Their wing emission components are extended up to $\gtrsim 500\kms$ from 
the line centers, and no clear counterparts are found on the 
blue side. The origin of such asymmetry, which is commonly seen in 
high-$z$ \lya\ emission, is thought to be either 
absorption by intervening neutral hydrogen (Lyman forest absorption), 
or galactic superwind. 
Although these two mechanisms cannot be definitely discriminated, 
the spectra of these five objects show some evidence 
that favor the existence of superwind activities as shown below. 

No significant continuum emission is detected in any of the 18 objects 
in our sample. Only one object, IB12-21989, showed marginal 
signature of continuum emission with $\sim 2\sigma$ level of the sky 
fluctuation, but this object does not have the red wing. 
This implies that the contribution of the continuum to the wing components 
we identified is negligible, and there should be a high velocity 
\lya\ emission component within the systems. 
An asymmetric line profile with a broad red wing and a sharp blue cut-off 
is commonly seen in superwind galaxies \citep[e.g.][]{ajiki02,dawson02}, 
and well agrees with theoretical predictions of the superwind model 
\citep{tenorio-tagle}. 

In fact, one of these five objects with red wings, IB10-90651, 
shows a systematic velocity 
structure on the two-dimensional spectrum (second part of Figure 
\ref{fig:spec}, top-left panel). 
This object has an extremely extended, diffuse emission component on 
the red side of the \lya. It extends up to $5''$ (projected distance of 
36 kpc at $z=3.68$) toward south from the center of the object, 
and the velocity extent is $\sim 1000\kms$ redward from the line center. 
On the blue side, although fairly faint, we can see a counter-image 
of the diffuse component extending toward north. 
These velocity structures suggest that this object has a galactic superwind 
flowing nearly along the position angle of the slit. 
The faintness of the blue component can be explained with 
absorption by the near-side shell of \ion{H}{1} gas, 
while the flux of the red component is enhanced by back-scattering 
by the far-side shell \citep{tenorio-tagle}. 

Although the blue counter-image can be seen in the two-dimensional 
spectrum of IB10-90651, we cannot see blue wing emission on the 
one-dimensional spectrum. 
This asymmetric profile is quite similar to other four objects. 
This suggests that all the five objects with red wings are 
likely to be superwinds. 
The fraction of such objects in our sample, $5/18\simeq 30\%$, 
is similar to that of sub-mm detection in M04's sample \citep{geach05}. 
Such sub-mm detection suggests the existence of obscured 
starburst regions responsible for extended \lya\ emission, 
i.e., superwinds. 
The lack of diffuse and extended high velocity components 
may suggest either that the direction of outflow is perpendicular 
to the slit direction, or that the \lya\ emission is intrinsically faint. 
The former possibility should be examined by further 
deep spectroscopy with various slit direction or an integral field 
spectroscopy. The latter is quite consistent with the 
relatively small EWs of the four objects with red wings. 
When the contribution of superwind activities is relatively small, 
the EWs of the \lya\ emission should be smaller than 
superwind-dominated sources like IB10-90651. 

Note that there are some other objects that possibly have 
red side wings, e.g. IB12-81981 ($\sim 1.5\sigma$ up to $\sim 400\kms$). 
However, the wing components of such objects, if any, are fairly 
faint, and affected by neighboring 
sky emission lines for most cases. If they indeed have wing emission 
on the red side, they are also likely to have superwind activities 
within the systems. Other notes for each object are listed in 
appendix \ref{notes:wing}

\subsection{Objects with large EWs and no wings}
\label{res:cool}
After classifying our objects in two ways, 
there remains eight objects that have large EWs 
and no significant wing emission on the red side of the \lya\ line: 
IB11-59167, IB11-80344, IB11-89537, IB12-30834, IB12-48320, IB12-58572, 
IB12-81981, and IB13-104299. 
Except for one, IB12-58572, they are all continuum-faint 
objects (see \S\ref{sample}) with $\rm EW_{rest} > 240\AA$. 
The EW of 240\AA\ is predicted for the starbursts with solar metallicity 
at an age of $10^6\rm yrs$ by \citet{cf93}, and also predicted by 
MR02 by introducing an extremely top-heavy IMF (slope = 0.5) 
or zero-metallicity, for starbursts with an age of $\gtrsim 10^7\rm yrs$. 
As mentioned above in \S\ref{spec:ew}, 
the observed EWs should be smaller than intrinsic values due to the 
absorption of \lya\ emission by the IGM. 
Namely, the seven continuum-faint objects 
have EWs securely exceeding 240\AA. 

For these eight sources, the origin of the \lya\ emission is hardly 
thought to be {\it  ordinary} starbursts. 
If their ionizing sources are starbursts, the large EWs require 
photo-ionization by extremely young stellar populations 
(age $\ll 10^7\rm yrs$) with an extremely top-heavy IMF 
and/or very low metallicity. 
In this case, the stellar components should be dominated by population 
III stars. Calculations by MR02 show that a stellar population 
with zero-metallicity and Salpeter's IMF can achieve an extremely 
large EW, 1122\AA\ at the age of $10^6\rm yrs$. \citet{schaerer03} 
also suggests by treating various IMFs and stellar mass ranges that, 
under very metal-poor conditions, the EWs of \lya\ emission 
from very young starbursts can be highly enhanced, $\sim 1300\rm\AA$ 
at an age of $10^6\rm yrs$. In this picture, the extended \lya\ 
emission can be attributed to resonant-scattering. 
In addition, it is thought that cooling radiation or supernova explosions 
also contribute to the \lya\ emission, at least in some part. 

As for the starbursts, it is also possible that the starburst 
regions are heavily obscured by dust with somewhat an inhomogeneous 
distribution. If the dust screen absorbs UV photons 
traveling along the line of sight, the observed faintness in UV continuum 
can be reproduced. Then if UV photons traveling along other 
directions are not significantly absorbed, they can ionize the 
\ion{H}{1} gas surrounding the starburst region, and can lead to 
spatially extended \lya\ emission. 
Since UV continuum emission cannot be scattered in this manner, 
these processes can at least qualitatively explain the extended 
\lya\ emission and faint / compact UV continuum emission. 
This is thought to be a plausible explanation because 
galaxies in high redshifts are generally surrounded by a 
larger amount of \ion{H}{1} gas than galaxies in the present 
universe \citep[e.g.][]{adelberger03}. 
The \lya\ emitting gas will be spatially more extended than 
the starburst region, and thus less sensitive to dust absorption. 

If the dust screen covers the whole direction, even \lya\ photons 
cannot escape the system. Instead, the star forming activities 
can lead to shock-ionization by a superwind 
\citep[e.g.][]{ts00,tsk01,veilleux05}. 
However, we could not detect significant high-velocity components 
on the \lya\ line profiles, nor on the two-dimensional spectra. 
If superwind activities are responsible for the extended \lya\ 
emission for these objects, the shocked region associated with 
the wind, if any, should be very faint. Their relatively small velocity 
widths of the \lya\ line, typically $\sim 300-500\kms$, also suggests 
that powerful superwinds are unlikely. The exception is two objects, 
IB11-80344 and IB11-89537, which have relatively large velocity widths 
up to $\sim 740\kms$. These objects might be superwinds, 
but have smaller velocity widths than the most prominent ones known 
to date \citep[e.g. $\sim 10^3\kms$, ][]{heckman90}. 
Additionally, note that there are some objects that have a {\em marginal} 
signature of wing emission on the red side of the \lya, 
e.g. IB12-48320 and IB12-81981. The red side spectra of these 
objects are overlapped with relatively strong sky emission lines 
which cannot be subtracted completely. If a deeper exposure and/or a 
more sufficient sky subtraction become available, we can possibly 
detect a high velocity wing component on the line profile for some 
objects. 

Apart from such uncertainties, the eight objects show a remarkable feature. 
Figure \ref{fig:ldv} plots the velocity width as a function of 
the \lya\ line luminosity. If the ionizing mechanism is 
cold accretion, the velocity width will roughly correspond to 
the circular velocity of the system, and thus will be scaled to 
the mass of the system. 
On the other hand, the \lya\ luminosity should be positively correlated 
with the gravitational energy of the system, and thus scaled to the mass 
of the system. Therefore, the velocity width and the \lya\ line luminosity 
should have a positive correlation. 
Among all the objects in our VIMOS sample, 
the eight objects with large EWs and no wings clearly show 
a positive correlation in Figure \ref{fig:ldv}. 
Although not a definite constraint, this result supports the 
idea of cooling radiation. 
Their relatively small velocity widths quite well agree with 
a theoretical prediction of cooling radiation \citep{fardal}. 
Even for the two objects with large velocity widths ($\Delta V>500\kms$), 
these velocity widths can be reproduced by taking into account 
the effect of resonant-scattering, or perhaps an additional ionizing 
source in the system \citep{dijkstra06a, dijkstra06b}. 
Furthermore, two or more objects (e.g. IB12-58572 and IB12-104299) 
marginally show an odd asymmetry of the line profile, i.e., broader 
on the blue side than on the red side, which also agrees with the 
prediction of \citet{dijkstra06a}. 
Thus, cooling radiation can be a reasonable explanation for these 
eight objects. It is also true that other mechanisms contribute
to the \lya\ emission, together with cold accretion. 
Such additional mechanisms are discussed in \S\ref{res:other}
Notes for individual objects are described in appendix \ref{notes:cool}.

\subsection{Constraints from other data}
\label{res:other}
We have mentioned about four possibilities for the origins of the 
spatially extended \lya\ emission: stellar photo-ionization, 
photo-ionization by AGNs, shock-heating by superwinds, 
and cooling radiation from infalling material (cold accretion). 
These scenarios predict characteristic emission lines of each 
ionizing source. 
In addition to our VIMOS data, we can use the data 
taken in our previous study and X-ray data for diagnostics 
of the physical origins. 
We then first analyzed the low-resolution spectroscopic 
data taken with FOCAS on Subaru Telescope, and the IA images 
taken with Suprime-Cam (Paper1) to detect the emission lines. 

For starbursts, there are two cases described in \S\ref{res:cool}, 
i.e., extremely young starbursts dominated by population III stars, 
and starbursts heavily obscured by dust. 
The former case of starbursts has an extremely low metallicity and 
a top-heavy IMF. Such an extremely young stellar population formed 
in a primordial condition should have \ion{He}{2} $\lambda 1640$ 
emission \citep{schaerer03}. 
The \ion{He}{2} line is also predicted for the case of cooling 
radiation \citep{yang06}. While \lya\ line is optically thick in 
general, this line should be optically thin. 
This implies that the \ion{He}{2} line is suitable to probe the 
gas dynamics, and thus a good probe to discriminate stellar photo-ionization 
and cooling radiation. However, none of our objects in the FOCAS 
spectroscopic sample has detectable \ion{He}{2} emission. 
On the composite spectrum, there is apparently an emission 
feature at the wavelength of rest-frame 1640\AA\ (see Figure 4 of 
Paper1). 
This is just a marginal signature since the wavelengths near the 
rest-frame 1640\AA\ is largely affected by sky emission lines. 

For the case of obscured starbursts, there should be characteristic 
metal lines, such as [\ion{O}{2}] $\lambda 3727$, [\ion{O}{3}] 
$\lambda\lambda 4959, 5007$, and H$\alpha$ $\lambda 6563$. 
These lines are less sensitive to dust absorption than the \lya\ line, 
and should be good diagnostics of starbursts, although they 
are redshifted beyond the wavelength coverage of our FOCAS data. 
Such metal lines can also be seen in superwind galaxies. 
The primordial gas surrounding galaxies should be 
chemically enriched by supernova explosions, resulting in 
spatially extended metal lines. Namely, superwind galaxies 
are expected to have spatially extended \ion{C}{4} line emission 
with a strength of $\sim 1/10$ of the \lya\ line \citep{heckman91b}. 
We previously showed in Paper1 that there are no such metal emission 
lines either in the individual spectra or in the stacked spectrum, 
down to our current detection limits. This suggests that at least 
the majority of the FOCAS sample (seven objects) are unlikely to 
be superwinds. 
These pieces of evidence from the observations are, however, just weak 
constraints on the superwind scenario. In fact, IB10-90651 is 
proved to have extended diffuse emission components by performing 
deep spectroscopy. Further deep follow-up observations may find 
diffuse emission components suggesting the superwind scenario. 

The \ion{C}{4} emission line, as well as \ion{N}{5} or \ion{He}{2} 
lines, can also be diagnostics of the AGN activities. 
Similarly to superwinds, AGN scenario is also likely to be ruled out 
by the absence of the \ion{C}{4} or \ion{N}{5} lines on the 
FOCAS spectra. 
Our IA images show that, although they are more extended than point-sources, 
the sizes of \lya\ emission components are typically $\sim 10-15\rm\, kpc$, 
and are smaller than those of \lya\ nebulae associated with AGNs 
\citep[e.g.][]{heckman91a,weidinger04,weidinger05}. 
The velocity widths of our objects are also smaller than those of 
\lya\ nebulae associated with AGNs known to date 
\citep[e.g.][]{vanojik,weidinger04,weidinger05}. 
Together with the faintness in UV continuum, 
these facts suggest that if the ionizing sources are AGNs, 
they must be obscured (type II) AGNs. 
In order to examine such AGN activities, it is important to compare 
them with type II QSOs known to date. 
Type II AGNs are, however, suggested to be quite rare by photometric 
studies of LAE samples \citep{malhotra03, wang04}. 
We will then just show the results of our analysis of X-ray data 
taken with {\it XMM-Newton}. 

Most of our field is covered by X-ray data with two pointings 
of {\it XMM-Newton} (Ueda et al. 2007), 
and the X-ray flux limits can be estimated by using these data. 
The sensitivity is sufficient to detect bright quasars at $z\sim 4$. 
In order to put the strongest constraints, 
we here used the data of 0.5--4.5 keV band, 
the most sensitive band of {\it XMM-Newton}. 
We estimated the $1\sigma$ flux limits in the rest-frame 2--10 keV 
from the count rate limits, by assuming X-ray spectra of 
$F_\epsilon\propto\epsilon^{-0.8}$ and absorption with a hydrogen 
column density of 
$N_{\rm H}=1\times 10^{23}\rm cm^{-2}$ and 
$N_{\rm H}=1\times 10^{24}\rm cm^{-2}$. 
The flux limits thus estimated are listed in Table \ref{tab:x}. 
The typical values of their flux limits in the case of 
$N_{\rm H}=1\times 10^{23}\rm cm^{-2}$ are a few $\times 10^{-15}\fluxunit$, 
which leads to X-ray luminosities $\sim 10^{45}\lum$ at $z\sim 4$. 
These are comparable to the X-ray luminosities of type II QSOs 
known to date \citep{norman02, stern02, dawson03}. 
Our analysis ignores the reflection components of X-ray emission, 
so that the upper limits listed in Table \ref{tab:x} are 
thought to be the most conservative values, i.e., the X-ray 
luminosities are likely to be significantly smaller than 
those of type II QSOs. 
Note that we cannot put any constraints if our objects are 
compton-thick, $N_{\rm H}\gg 10^{24}\rm cm^{-2}$. 
These imply that our objects are unlikely to be harboring 
type II QSOs, although the possibility of AGNs is not completely 
ruled out due to the relatively shallow exposure of the 
X-ray data.

\section{Conclusions}
\label{con}
We have carried out deep, medium-resolution ($R\sim 2000$) follow-up 
spectroscopy of 18 extended \lya\ sources at $z\sim 3-5$ that are faint 
and/or compact in UV continuum, using VLT/VIMOS. 
We found that all 18 objects in our VIMOS sample have fairly large 
equivalent widths (EWs) of the \lya\ emission with a median value 
of $\approx 210\rm\AA$ in the rest-frame. 
A half of our sample (nine objects) have moderately large EWs of 
$\sim 100-200\rm\AA$, and are accounted for by ordinary starbursts with 
a solar metallicity. The high resolution and S/N ratio of our data 
enabled us to make a quantitative analysis of the \lya\ line profiles. 
The velocity widths were found to be relatively small, 
typically $\sim 300-500\kms$. 

For five objects ($\sim 30\%$), we identified conspicuous 
broad wing emission components on the red side of the \lya\ line, 
while no significant wing emission was found on the blue side. 
Such features quite well agree with theoretical predictions of 
galactic superwind model. 
Namely, we found diffuse high velocity emission components extended 
up to $\sim 70\rm\, kpc$ on the two-dimensional spectrum of 
one of these five objects. 
The velocity extent of these components is $\gtrsim 2000\kms$ 
with systematic velocity structure, i.e., the southern part 
is redshifted and the northern part is blueshifted. 
These features suggest the existence of galactic superwind activities. 

Excluding the objects above, there remain eight objects, which 
have large EWs exceeding 200\AA\ (seven have EWs exceeding 240\AA), 
and no clear signature of high velocity wing emission on the either 
side of the \lya\ line. These objects are hardly thought to be 
ordinary starbursts, and our IA images and FOCAS spectra show 
no clear signature of AGN activities. Non-detection in X-ray 
data taken with {\it XMM-Newton} also suggests they are unlikely to 
be AGN-origin like type II QSOs. Their relatively small spatial 
extents and velocity widths agree well with theoretical predictions 
of cooling radiation. Furthermore, their velocity widths clearly 
show a positive correlation with the \lya\ line luminosities. 
These facts suggest that they are candidates for forming-galaxies 
in gas-cooling phase, i.e., the very first stage of galaxy formation.

\appendix
\section{Notes for individual objects}
\subsection{Objects with moderate EWs}
\label{notes:modEW}
\paragraph{IB08-86220}
$z=3.31$, ${\rm SFR} \simeq 3.3\mdot$, $\rm EW_{rest}\simeq 120\rm\AA$. 
Even the upper limit of EW does not exceed 200\AA. 
Our analysis may suffer from relatively large residuals of the 
sky emission at $\approx 5225\rm\AA$ and $\approx 5240\rm\AA$. 

\paragraph{IB10-17108}
$z=3.79$, ${\rm SFR}\simeq 10\mdot$, $\rm EW_{rest}\simeq 110\rm\AA$. 
The EW is the smallest in our sample, and the uncertainty in the 
EW is relatively small. Wing emission can be seen on the red side 
(see appendix \ref{notes:wing}). 

\paragraph{IB10-32162}
$z=3.74$, ${\rm SFR}\simeq 9.7\mdot$, $\rm EW_{rest}\simeq 150\rm\AA$. 
The upper limit of the EW is relatively large for this group of objects, 
$\simeq 180\rm\AA$. 
The velocity width is relatively large, $\Delta V \simeq 610\kms$ (FWHM). 
Wing emission can be seen on the red side (see appendix \ref{notes:wing}). 

\paragraph{IB10-54185}
$z=3.82$, ${\rm SFR}\simeq 6.5\mdot$, $\rm EW_{rest}\simeq 130\rm\AA$. 
The velocity width is quite small, $\Delta V\simeq 300\kms$. 
Wing emission can be seen on the red side (see appendix \ref{notes:wing}). 

\paragraph{IB12-21989}
$z=4.11$, ${\rm SFR}\simeq 13\mdot$, $\rm EW_{rest}\simeq 180\rm\AA$. 
Continuum emission can be seen at $\simeq 1.5\sigma$ level, 
and the EW calculated directly from the spectrum is $\simeq 35\rm\AA$. 
Our photometry may suffer from neighboring continuum sources. 

\paragraph{IB12-71781}
$z=4.11$, ${\rm SFR}\simeq 4.1\mdot$, $\rm EW_{rest}\simeq 130\rm\AA$. 
The upper limit of the EW is fairly large, $\simeq 230\rm\AA$, 
so that this object is possibly not categorized as an ordinary starburst. 
Our analysis may suffer from the several sky emission lines on the 
red side of the \lya\ (e.g. $\approx 6235\rm\AA$). 

\paragraph{IB13-96047}
$z=4.27$, ${\rm SFR}\simeq 10\mdot$, $\rm EW_{rest}\simeq 140\rm\AA$. 
Even the upper limit of the EW does not exceed 200\AA. 
Our spectral analysis may suffer from neighboring several sky 
emission lines on the both sides of the \lya. 
Neighboring continuum sources may affect our photometric analysis. 

\paragraph{IB14-47257}
$z=4.66$, ${\rm SFR}\simeq 13\mdot$, $\rm EW_{rest}\simeq 180\rm\AA$. 
With an upper limit of the EW $\simeq 240\rm\AA$, this object not be 
an ordinary starburst. The spectrum suffers from relatively strong residuals of 
the sky emission at $\approx 6865, 6900, \mbox{and } 6910\rm\AA$ near 
the \lya\ line. The \lya\ line itself is overlapped with weak sky emissions. 

\paragraph{IB14-52102}
$z=4.47$, ${\rm SFR}\simeq 9.0\mdot$, $\rm EW_{rest}\simeq 150\rm\AA$. 
With an upper limit of the EW $\simeq 200\rm\AA$, this object may not 
be an ordinary starburst. The line profile has wing emission on the red side 
(see appendix \ref{notes:wing}). 

\subsection{Objects with broad asymmetric wings}
\label{notes:wing}
\paragraph{IB10-17108}
$z=4.11$, $\rm EW_{rest}\approx130\rm\AA$, $\Delta V \approx490\kms$. 
Also listed in appendix \ref{notes:modEW}. 
The wing component is very clear, and extended up to $\gtrsim 700\kms$ 
from the line center. 

\paragraph{IB10-32162}
$z=3.74$, $\rm EW_{rest}\approx150\rm\AA$, $\Delta V \approx610\kms$. 
Also listed in appendix \ref{notes:modEW}. 
The velocity width of the main component is fairly large, $\approx610\kms$. 
The wing component is not very clear, but the line profile shows 
a significant asymmetry. 

\paragraph{IB10-54185}
$z=3.82$, $\rm EW_{rest}\approx130\rm\AA$, $\Delta V \approx300\kms$. 
Also listed in appendix \ref{notes:modEW}. 
The spectrum may suffer from sky emission lines on the red side 
($\approx 5890\rm\AA$ and a weak one at $\approx 5870\rm\AA$), 
but the red side wing exceeds the $3\sigma$ noise level at $\lesssim 450\kms$ 
from the line center. The wing component can be seen up to 
$\sim 600-1000\kms$. 

\paragraph{IB10-90651}
$z=3.68$, $\rm EW_{rest}\approx860\rm\AA$, $\Delta V \approx570$. 
The EW is the largest in our sample. 
The wing emission is extended up to $\gtrsim 500\kms$. 
A systematic velocity structure of diffuse emission can be seen 
in the two-dimensional spectrum. 

\paragraph{IB14-52102}
$z=4.47$, $\rm EW_{rest}\approx150\rm\AA$, $\Delta V \approx350\kms$. 
Also listed in appendix \ref{notes:modEW}. 
The wing component exceeds the $3\sigma$ noise level, and is extended 
up to $\sim 500\kms$.

\subsection{Objects with large EWs and no wings}
\label{notes:cool}
\paragraph{IB11-59167}
$z=3.94$, $\rm EW_{rest}\approx 270\rm\AA$, $\Delta V \approx270 \kms$. 
A systematic (outflow-like) velocity structure may exist, 
but the spectrum may suffer from incomplete sky subtraction 
at $\approx 6000\rm\AA$. 

\paragraph{IB11-80344}
$z=3.89$, $\rm EW_{rest}\approx 569\rm\AA$, $\Delta V\approx740\kms$. 
The lower limit of the EW is $\approx 380\rm\AA$. 
The velocity width is large, and the line profile has a double-peaked 
shape. The spectrum may suffer from several sky emission lines. 

\paragraph{IB11-89537}
$z=4.03$, $\rm EW_{rest}\approx 460\rm\AA$, $\Delta V\approx 560\kms$. 
The lower limit of the EW is $\approx 300\rm\AA$. 
Wing emission on the red side may exist. 
The spectrum may suffer from a neighboring continuum source. 

\paragraph{IB12-30834}
$z=4.11$, $\rm EW_{rest}\approx 360\rm\AA$, $\Delta V\approx 430\kms$. 
The lower-limit of EW is $\approx 240\rm\AA$. 
The line profile is quite simple (nearly Gaussian), and the 
velocity width is relatively small. 

\paragraph{IB12-48320}
$z=4.04$, $\rm EW_{rest}\approx 320$, $\Delta V\approx 320\kms$. 
The lower limit of the EW is $\approx 200\rm\AA$. 
A relatively strong residual of sky subtraction remains 
on the red side at $\approx 6145\rm\AA$. 
The line profile may suffer from a weak sky emission line here. 

\paragraph{IB12-58572}
$z=4.04$, $\rm EW_{rest}\approx 210\rm\AA$, $\Delta V\approx 400\kms$. 
The lower limit of the EW does not exceed 200\AA. 
This object was categorized as a continuum-compact source in 
appendix \ref{sample}. 

\paragraph{IB12-81981}
$z=4.12$, $\rm EW_{rest}\approx 500\rm\AA$, $\Delta V\approx 380\kms$. 
The lower limit of the EW is $\approx 260\rm\AA$. 
The profile shows asymmetry, but no significant wing emission can be 
found. The red side suffers from relatively strong sky emission lines 
at $\approx 6235\rm\AA$ and 6260\AA, and several weak ones. 

\paragraph{IB13-104299}
$z=4.42$, $\rm EW_{rest}\approx 300\rm\AA$, $\Delta V\approx 260\kms$. 
The lower-limit of EW does not exceed 200\AA. 
The spectrum may suffer from several sky emission lines 
at $\approx 6560, 6580, \mbox{ and }6600\rm\AA$. 
The photometry may suffer from a neighboring continuum source.


\begin{deluxetable}{cccccccc}
\tablecaption{Photometric properties of the VIMOS spectroscopic sample
\label{tab:phot}}
\tabletypesize{\scriptsize}
\tablewidth{0pt}
\tablehead{
\colhead{}&
\colhead{$\alpha$}&
\colhead{$\delta$}&
\multicolumn{3}{c}{\lya}&
\multicolumn{2}{c}{Continuum}\\
\colhead{Object ID}&
\multicolumn{2}{c}{(J2000)\tablenotemark{a}}&
\colhead{band}&
\colhead{mag($2''$)}&
\colhead{mag(auto)}&
\colhead{band}&
\colhead{mag($2''$)}
}
\startdata
IB08-86220&  2:18:28.34& -5:18:12.2& IA527& 25.76& 25.40& $R$& 
26.51\tablenotemark{c}\\
IB10-17108&  2:16:58.05& -5:34:19.2& IA574& 25.07& 24.83& $R$& 
25.65\tablenotemark{c}\\
IB10-32162&  2:16:56.90& -5:30:29.7& IA574& 25.22& 24.90& $R$& 
25.98\tablenotemark{c}\\
IB10-54185&  2:17:59.49& -5:25:07.5& IA574& 25.61& 25.13& $R$& 
26.44\tablenotemark{c}\\
IB10-90651&  2:17:43.35& -5:16:12.4& IA574& 25.84& 25.50& $R$& 
27.75\tablenotemark{f}\\
IB11-59167&  2:17:10.21& -5:23:47.5& IA598& 25.80& 25.54& $i'$& 
27.59\tablenotemark{f}\\
IB11-80344&  2:17:44.7&  -5:18:15.3& IA598& 25.53& 25.24& $i'$& 
27.03\tablenotemark{f}\\
IB11-89537&  2:17:45.3&  -5:15:53.6& IA598& 25.80& 25.52& $i'$& 
27.59\tablenotemark{f}\\
IB12-21989&  2:17:13.35& -5:32:56.6& IA624& 25.71& 25.06& $i'$& 
25.88\tablenotemark{c}\\
IB12-30834&  2:17:55.98& -5:30:53.4& IA624& 25.82& 25.71& $i'$& 
27.09\tablenotemark{f}\\
IB12-48320&  2:16:55.89& -5:26:37.1& IA624& 25.82& 25.35& $i'$& 
27.35\tablenotemark{f}\\
IB12-58572&  2:17:49.10& -5:24:11.6& IA624& 25.97& 25.32& $i'$& 
26.82\tablenotemark{c}\\
IB12-71781&  2:17:38.96& -5:20:58.1& IA624& 25.96& 25.67& $i'$& 
26.77\tablenotemark{c}\\
IB12-81981&  2:18:14.72& -5:18:32.5& IA624& 25.65& 25.27& $i'$& 
27.84\tablenotemark{f}\\
IB13-96047&  2:18:13.33& -5:15:05.4& IA651& 25.80& 25.16& $i'$& 
25.97\tablenotemark{c}\\
IB13-104299& 2:18:21.09& -5:13:25.2& IA651& 25.87& 25.48& $i'$& 
27.45\tablenotemark{f}\\
IB14-47257&  2:18: 1.96& -5:25:25.3& IA679& 25.93& 25.14& $i'$& 
26.29\tablenotemark{c}\\
IB14-52102&  2:18:00.12& -5:24:10.6& IA679& 25.39& 25.15& $i'$& 
26.33\tablenotemark{c}
\enddata
\tablenotetext{a}{Coordinates are based on SXDS version 1 astrometry.}
\tablenotetext{c}{Continuum-compact objects: 
objects which can be regarded as point-sources.}
\tablenotetext{f}{Continuum-faint objects: 
objects which are fainter than $3\sigma$ in continuum.}
\end{deluxetable}


\begin{deluxetable}{cccccccc}
\tablecaption{Spectral properties\label{tab:spec}}
\tabletypesize{\scriptsize}
\tablewidth{0pt}
\tablehead{
\colhead{Object ID}&
\colhead{$\lambda_c$\tablenotemark{a}}&
\colhead{$z$}&
\colhead{$F(\lya)$\tablenotemark{b}}&
\colhead{$L(\lya)$}&
\colhead{$\Delta V$ (FWHM)}&
\colhead{$EW_{rest}$}&
\colhead{wing}\\
\colhead{}&
\colhead{[\AA]}&
\colhead{}&
\colhead{[$10^{-17}\fluxunit$]}&
\colhead{[$10^{42}\lum$]}&
\colhead{[\kms]}&
\colhead{[\AA]}&
\colhead{}
}
\startdata
IB08-86220& 5234& 3.31& $3.3(1.9)\pm 0.3$& $3.3\pm 0.5$& 
$423^{+55}_{-20}$& $121^{+34}_{-25}$& -\\
IB10-17108& 5824& 3.79& $7.6(4.2)\pm 0.7$& $10\pm 2$& 
$492^{+60}_{-21}$& $112^{+19}_{-17}$& red\\
IB10-32162& 5761& 3.74& $7.4(3.6)\pm 0.6$& $9.7\pm 1.7$& 
$614^{+55}_{-31}$& $149^{+29}_{-24}$& red\\
IB10-54185& 5864& 3.82& $4.2(2.1)\pm 0.4$& $6.5\pm 1.3$& 
$295^{+21}_{-53}$& $128^{+35}_{-27}$& red\\
IB10-90651& 5691& 3.68& $8.3(4.4)\pm 1.9$& $11\pm 2$& 
$573^{+44}_{-176}$& $867^{+1110}_{-411}$& red\\
IB11-59167& 6006& 3.94& $2.2(1.2)\pm 0.3$& $3.4\pm 0.8$& 
$273^{+15}_{-47}$& $265^{+434}_{-117}$& -\\
IB11-80344& 5950& 3.89& $7.9(3.5)\pm 0.8$& $11\pm 3$& 
$736^{+63}_{-60}$& $569^{+387}_{-189}$& -\\
IB11-89537& 6115& 4.03& $5.3(2.9)\pm 0.5$& $8.4\pm 1.4$& 
$555^{+88}_{-50}$& $458^{+402}_{-164}$& -\\
IB12-21989& 6212& 4.11& $7.7(3.7)\pm 0.7$& $13\pm 2$& 
$454^{+56}_{-97}$& $184^{+43}_{-34}$& -\\
IB12-30834& 6217& 4.11& $4.9(2.6)\pm 0.5$& $8.0\pm 1.4$& 
$434^{+28}_{-44}$& $357^{+255}_{-119}$& -\\
IB12-48320& 6124& 4.04& $3.4(1.5)\pm 0.4$& $5.3\pm 1.2$& 
$320^{+54}_{-80}$& $318^{+333}_{-123}$& -\\
IB12-58572& 6129& 4.04& $3.6(1.3)\pm 0.5$& $5.7\pm 2.0$& 
$398^{+60}_{-22}$& $208^{+118}_{-66}$& -\\
IB12-71781& 6208& 4.11& $2.5(1.4)\pm 0.6$& $4.1\pm 1.8$& 
$291^{+67}_{-61}$& $134^{+95}_{-55}$& -\\
IB12-81981& 6224& 4.12& $3.4(1.9)\pm 0.4$& $5.7\pm 1.1$& 
$379^{+32}_{-56}$& $500^{+1497}_{-240}$& -\\
IB13-96047& 6592& 4.27& $5.4(2.0)\pm 0.5$& $10\pm 3$& 
$354^{+39}_{-57}$& $135^{+34}_{-26}$& -\\
IB13-104299& 6410& 4.42& $3.2(2.0)\pm 0.4$& $5.7\pm 1.2$& 
$259^{+23}_{-16}$& $303^{+393}_{-129}$& -\\
IB14-47257& 6878& 4.66& $5.6(2.0)\pm 0.8$& $13\pm 5$& 
$325^{+84}_{-31}$& $177^{+66}_{-47}$& -\\
IB14-52102& 6647& 4.47& $4.5(2.5)\pm 0.3$& $9.0\pm 1.2$& 
$353^{+56}_{-36}$& $152^{+47}_{-33}$& red
\enddata
\tablenotetext{a}{Wavelength of the line center.}
\tablenotetext{b}{The values before slit-loss-correction are shown in 
the parenthesises.}
\end{deluxetable}


\begin{deluxetable}{ccccccc}
\tablecaption{$1\sigma$ flux limits in X-ray\label{tab:x}}
\tabletypesize{\scriptsize}
\tablewidth{0pt}
\tablehead{
\colhead{} &
\colhead{} &
\colhead{} &
\multicolumn{2}{c}{$(N_{\rm H}=10^{23}\rm cm^{-2})$}&
\multicolumn{2}{c}{$(N_{\rm H}=10^{24}\rm cm^{-2})$}\\
\colhead{Object ID}&
\colhead{$z$}&
\colhead{C.R.L\tablenotemark{a}} &
\colhead{$F_{2-10}$\tablenotemark{b}} &
\colhead{$L_{2-10}$\tablenotemark{c}} &
\colhead{$F_{2-10}$\tablenotemark{b}} &
\colhead{$L_{2-10}$\tablenotemark{c}}\\
\colhead{}&
\colhead{}&
\colhead{[counts s$^{-1}$]}&
\colhead{[\fluxunit]}&
\colhead{[\lum]} &
\colhead{[\fluxunit]}&
\colhead{[\lum]}
}
\startdata
IB08-86220& 3.31& $4.42\times 10^{-4}$& $1.18\times 10^{-15}$& 
$1.45\times 10^{45}$& $4.73\times 10^{-15}$& $5.83\times 10^{45}$\\
IB10-17108& 3.79& $1.40\times 10^{-3}$& $3.30\times 10^{-15}$& 
$5.62\times 10^{45}$& $1.19\times 10^{-14}$& $2.03\times 10^{46}$\\
IB10-32162& 3.74& $6.3\times 10^{-4}$& $1.51\times 10^{-15}$& 
$2.49\times 10^{45}$& $5.51\times 10^{-15}$& $9.09\times 10^{45}$\\
IB10-54185& 3.82& $6.67\times 10^{-4}$& $1.57\times 10^{-15}$& 
$2.72\times 10^{45}$& $5.63\times 10^{-15}$& $9.76\times 10^{45}$\\
IB10-90651& 3.68& $4.46\times 10^{-4}$& $1.08\times 10^{-15}$& 
$1.71\times 10^{45}$& $4.00\times 10^{-15}$& $6.37\times 10^{45}$\\
IB11-59167& 3.94& $3.31\times 10^{-4}$& $7.63\times 10^{-16}$& 
$1.42\times 10^{45}$& $2.66\times 10^{-15}$& $4.96\times 10^{45}$\\
IB11-80344& 3.89& $4.51\times 10^{-4}$& $1.05\times 10^{-15}$& 
$1.90\times 10^{45}$& $3.70\times 10^{-15}$& $6.70\times 10^{45}$\\
IB11-89537& 4.03& $4.43\times 10^{-4}$& $1.00\times 10^{-15}$& 
$1.97\times 10^{45}$& $3.43\times 10^{-15}$& $6.75\times 10^{45}$\\
IB12-21989& 4.11& $7.55\times 10^{-4}$& $1.68\times 10^{-15}$& 
$3.47\times 10^{45}$& $5.67\times 10^{-15}$& $1.17\times 10^{46}$\\
IB12-30834& 4.11& $1.60\times 10^{-3}$& $3.58\times 10^{-15}$& 
$7.38\times 10^{45}$& $1.21\times 10^{-14}$& $2.50\times 10^{46}$\\
IB12-48320& 4.04& $5.97\times 10^{-4}$& $1.35\times 10^{-15}$& 
$2.67\times 10^{45}$& $4.61\times 10^{-15}$& $9.13\times 10^{45}$\\
IB12-58572& 4.04& $5.27\times 10^{-4}$& $1.19\times 10^{-15}$& 
$2.35\times 10^{45}$& $4.07\times 10^{-15}$& $8.06\times 10^{45}$\\
IB12-71781& 4.11& $3.98\times 10^{-4}$& $8.89\times 10^{-16}$& 
$1.83\times 10^{45}$& $2.99\times 10^{-15}$& $6.16\times 10^{45}$\\
IB12-81981& 4.12& $6.28\times 10^{-4}$& $1.40\times 10^{-15}$& 
$2.90\times 10^{45}$& $4.70\times 10^{-15}$& $9.75\times 10^{45}$\\
IB13-96047& 4.27& $7.15\times 10^{-4}$& $1.55\times 10^{-15}$& 
$3.50\times 10^{45}$& $5.06\times 10^{-15}$& $1.14\times 10^{46}$\\
IB13-104299& 4.42& $6.47\times 10^{-4}$& $1.37\times 10^{-15}$& 
$3.36\times 10^{45}$& $4.34\times 10^{-15}$& $1.06\times 10^{46}$\\
IB14-47257& 4.66& $7.18\times 10^{-4}$& $1.48\times 10^{-15}$& 
$3.80\times 10^{45}$& $4.52\times 10^{-15}$& $1.25\times 10^{46}$\\
IB14-52102& 4.47& $6.6\times 10^{-4}$& $1.39\times 10^{-15}$& 
$3.49\times 10^{45}$& $4.35\times 10^{-15}$& $1.09\times 10^{46}$\\
\enddata
\tablenotetext{a}{Count rate limits in the observed-frame 0.5--4.5 keV.}
\tablenotetext{b}{Flux limits in the rest-frame 2--10 keV.}
\tablenotetext{c}{Luminosity limits in the rest-frame 2--10 keV.}

\end{deluxetable}


\begin{figure}
\figurenum{4}
\plotone{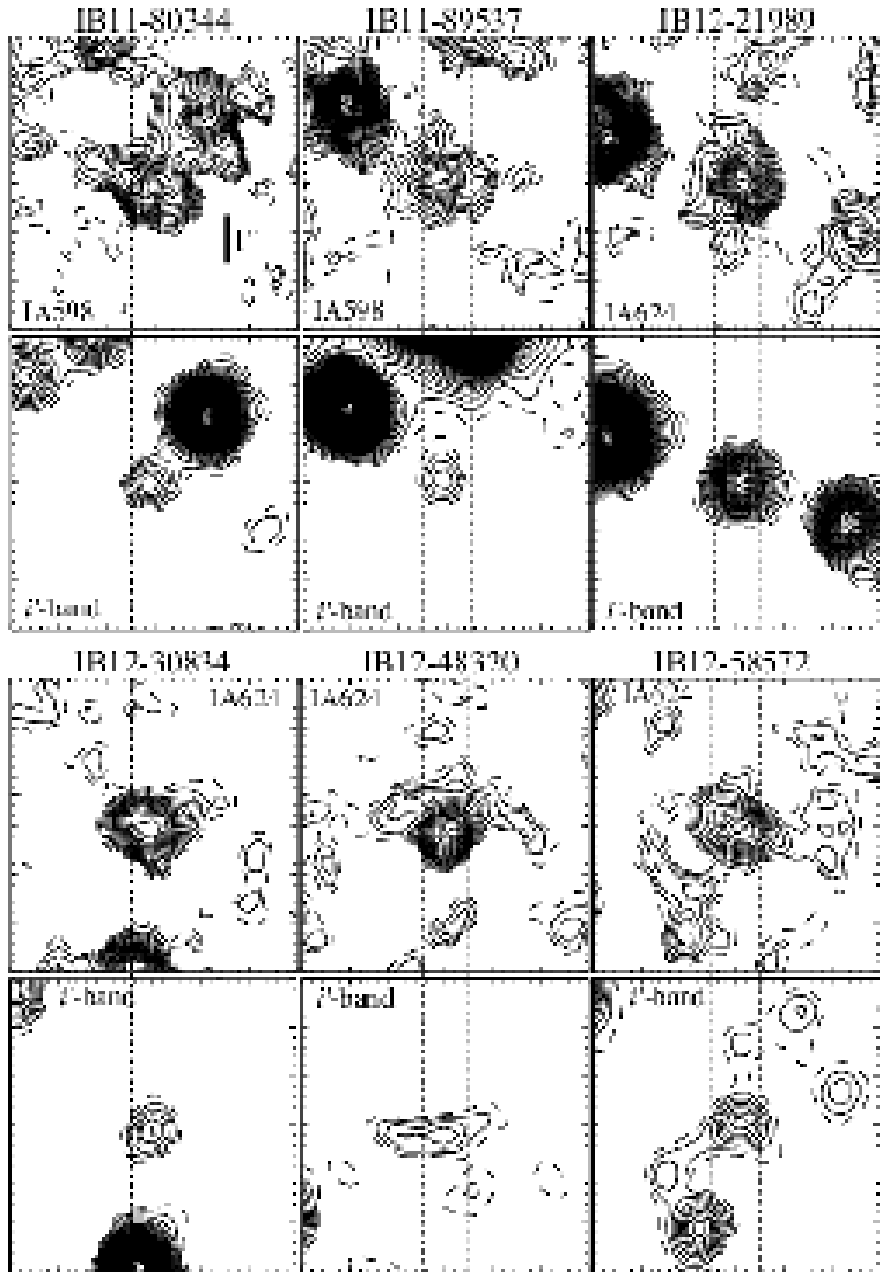}
\caption{Images of the 18 extended \lya\ sources. (Continued)}
\end{figure}

\begin{figure}
\figurenum{4}
\plotone{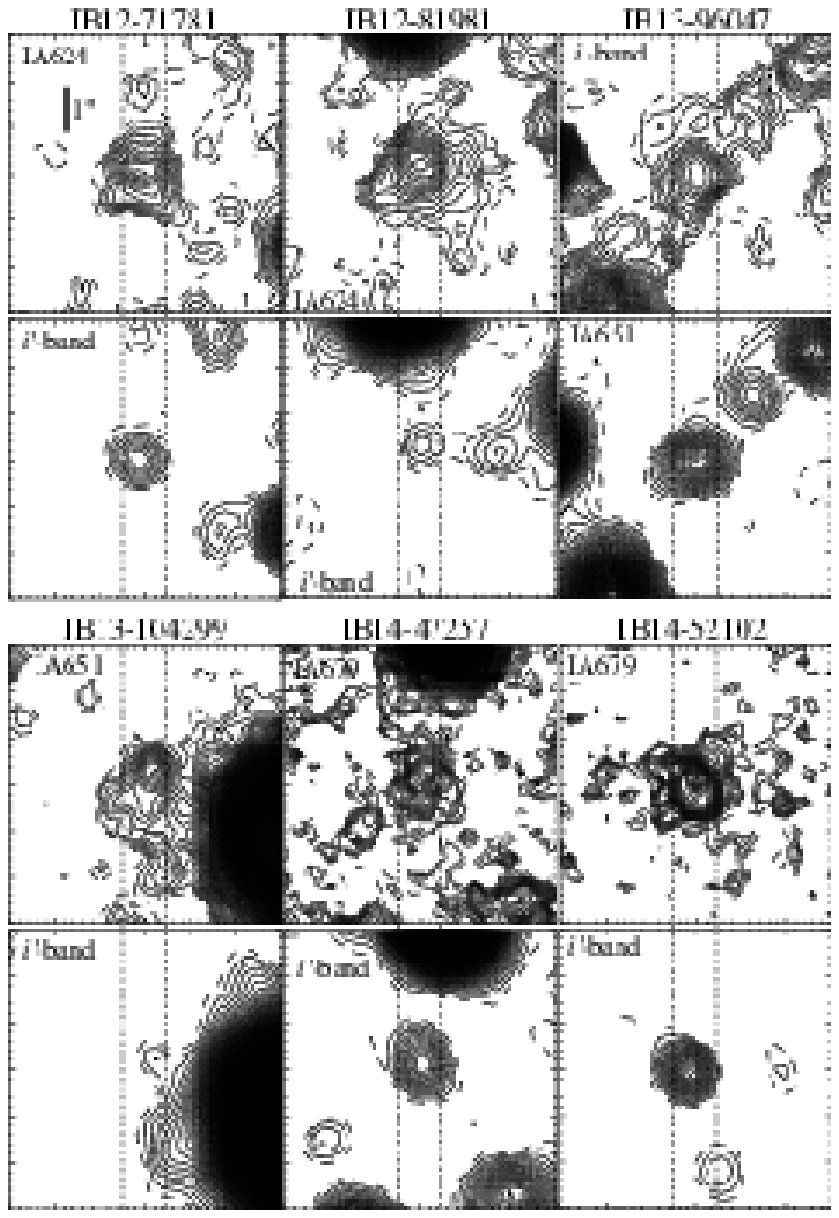}
\caption{Images of the 18 extended \lya\ sources. (Continued)}
\end{figure}

\begin{figure}
\plotone{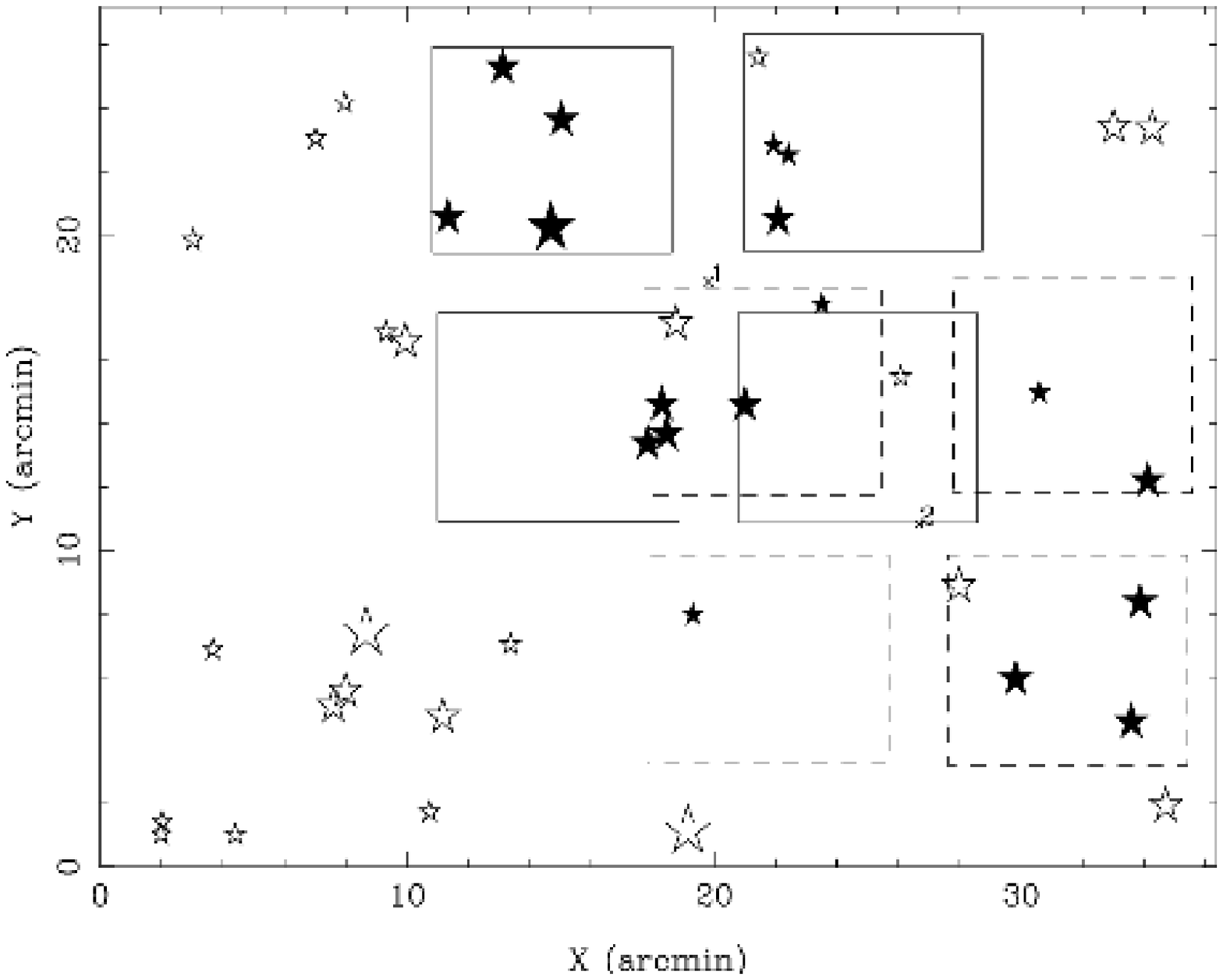}
\caption{Sky distribution of our objects. 
The solid and dashed rectangles show the two MOS fields 
of the VIMOS spectroscopy. The 41 objects in the photometric sample are 
shown in stars, and the 18 objects in the VIMOS sample are 
shown in filled symbols. 
The large symbols represent objects brighter than 25.5 mag (AB) 
in the IA bands with large IA-excess (Cont - IA $>$ 2.0 mag) 
The medium large symbols are objects brighter than 25.5 mag in the 
IA bands. }
\label{fig:skydist}
\end{figure}

\begin{figure}
\plotone{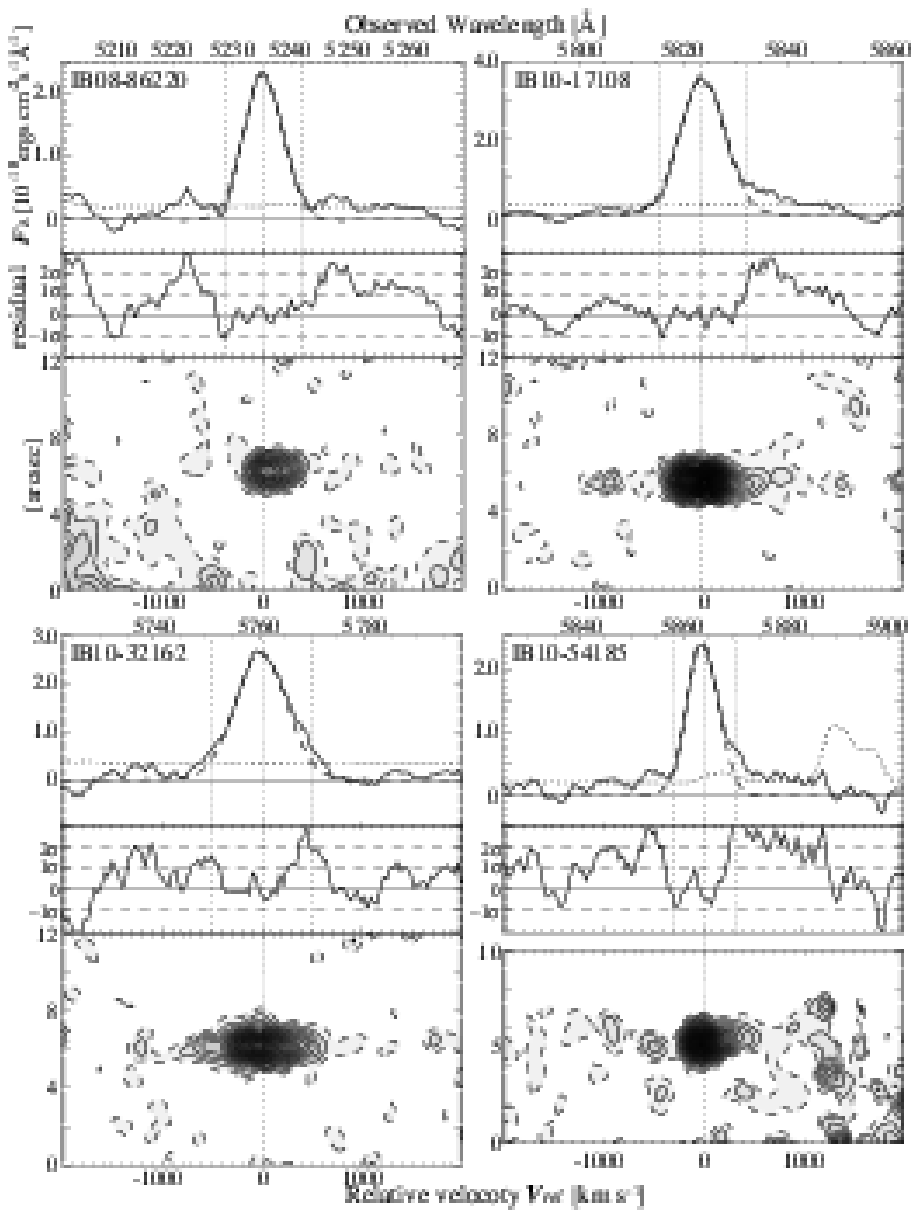}
\caption{Spectra taken with VIMOS. For each object, three panels 
show line profile, residual of Gaussian fitting, and two-dimensional 
spectrum (from top to bottom). 
On the top panel, the solid curve shows line profile, the dotted 
curve shows sky spectrum scaled to 1/100, and the vertical dotted 
lines show central wavelength and $\pm 1\sigma$ of the Gaussian 
function fitted to the line profile. The Gaussian function is shown with 
the dashed curve. }
\label{fig:spec}
\end{figure}

\begin{figure}
\figurenum{6}
\plotone{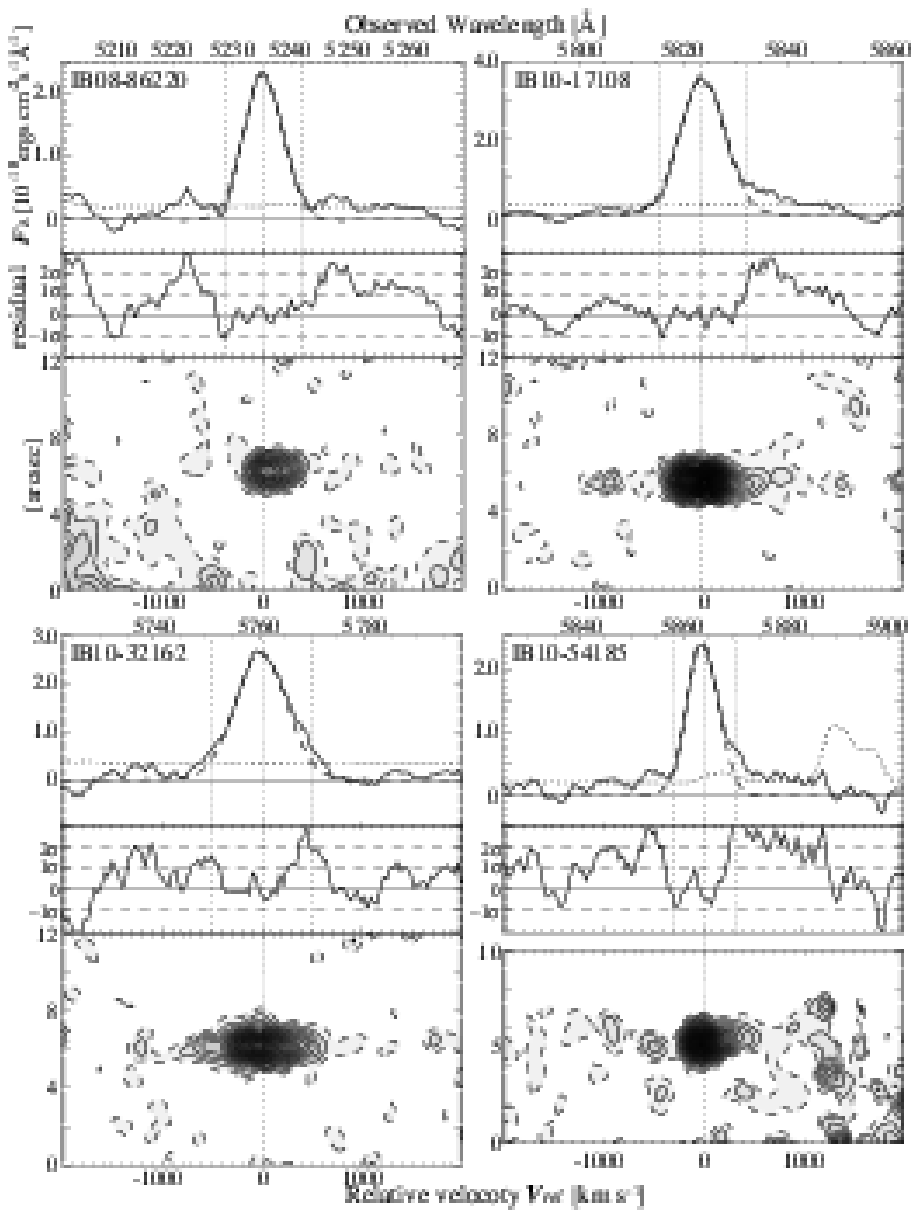}
\caption{Spectra taken with VIMOS. (Continued)}
\end{figure}

\begin{figure}
\figurenum{6}
\plotone{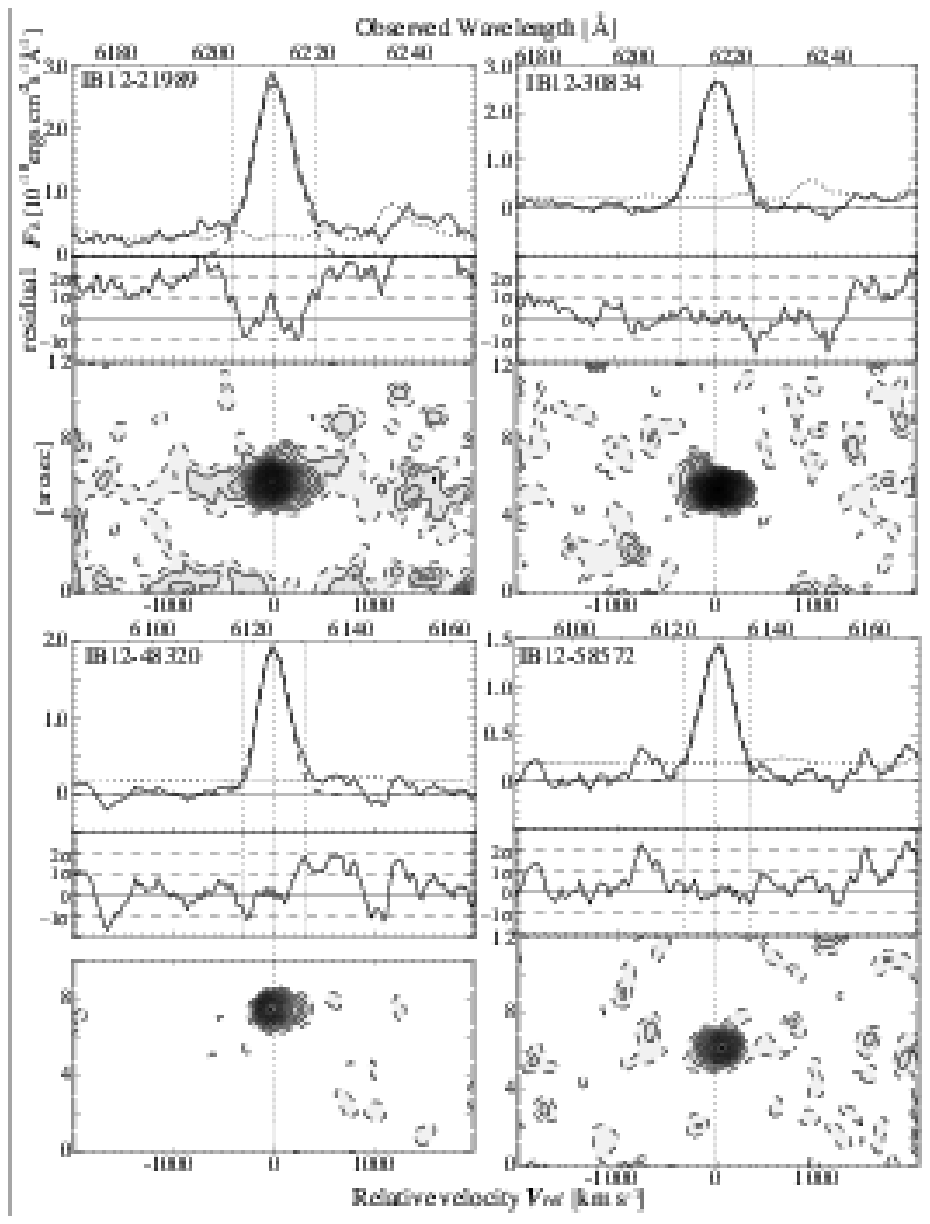}
\caption{Spectra taken with VIMOS. (Continued)}
\end{figure}

\begin{figure}
\figurenum{6}
\plotone{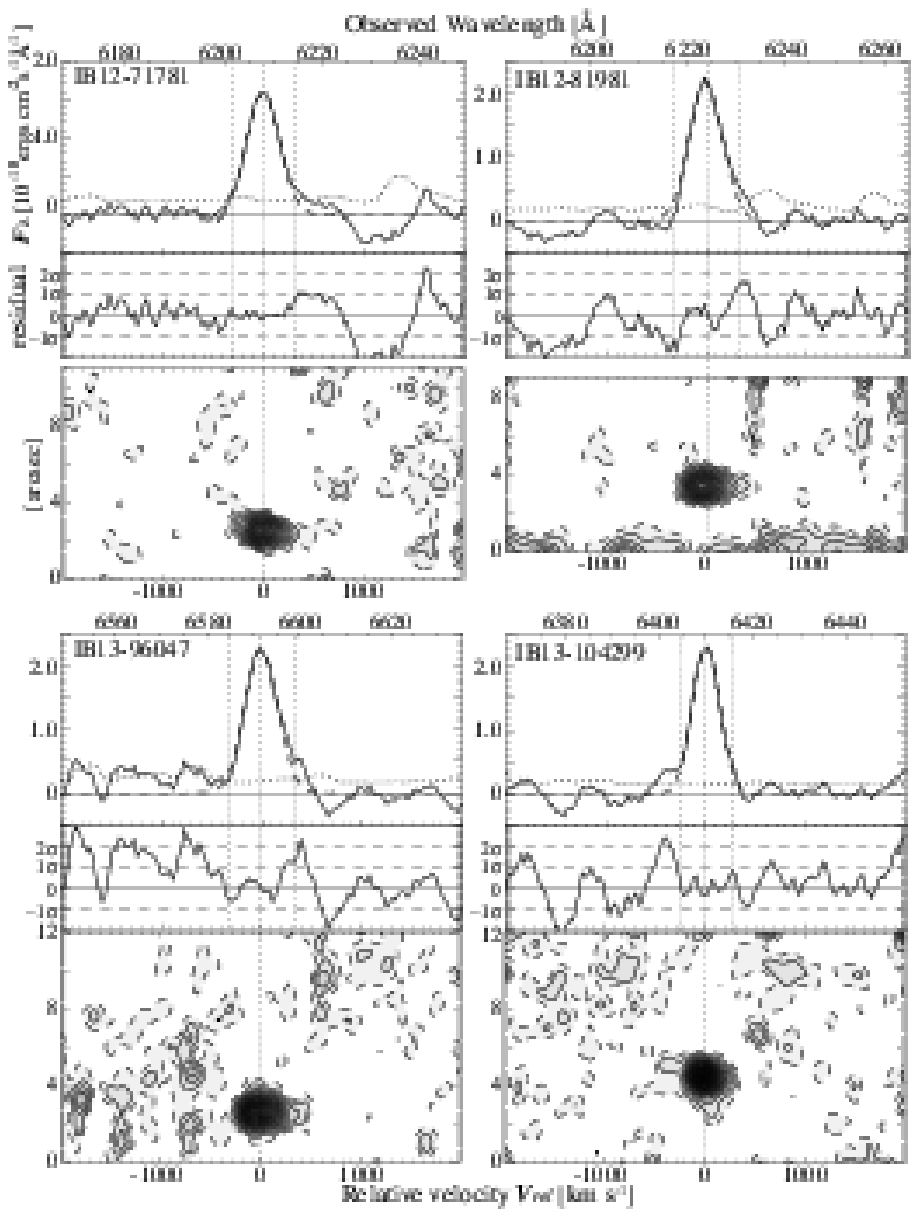}
\caption{Spectra taken with VIMOS (Continued)}
\end{figure}

\begin{figure}
\figurenum{6}
\plotone{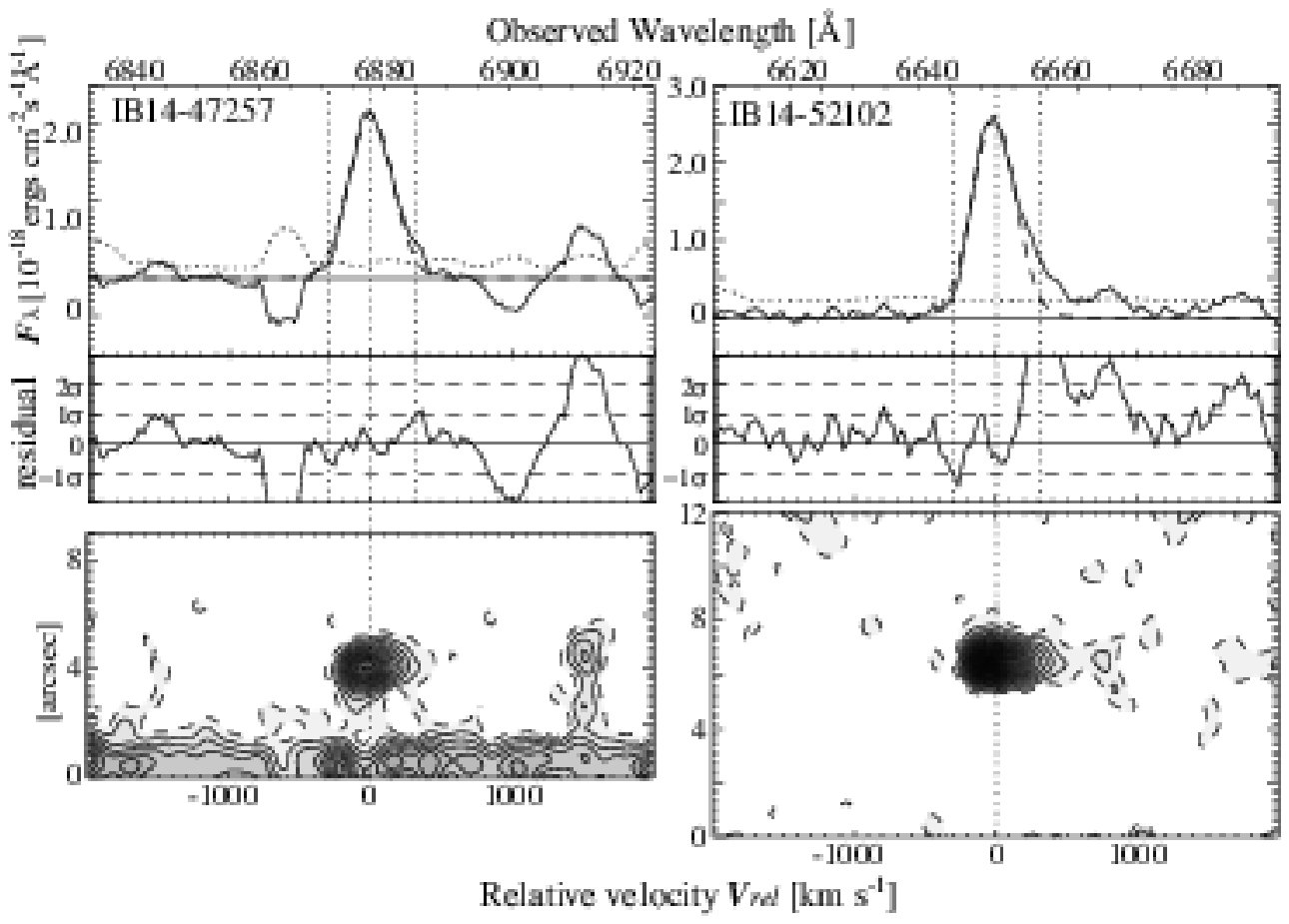}
\caption{Spectra taken with VIMOS (Continued)}
\end{figure}

\begin{figure}
\plotone{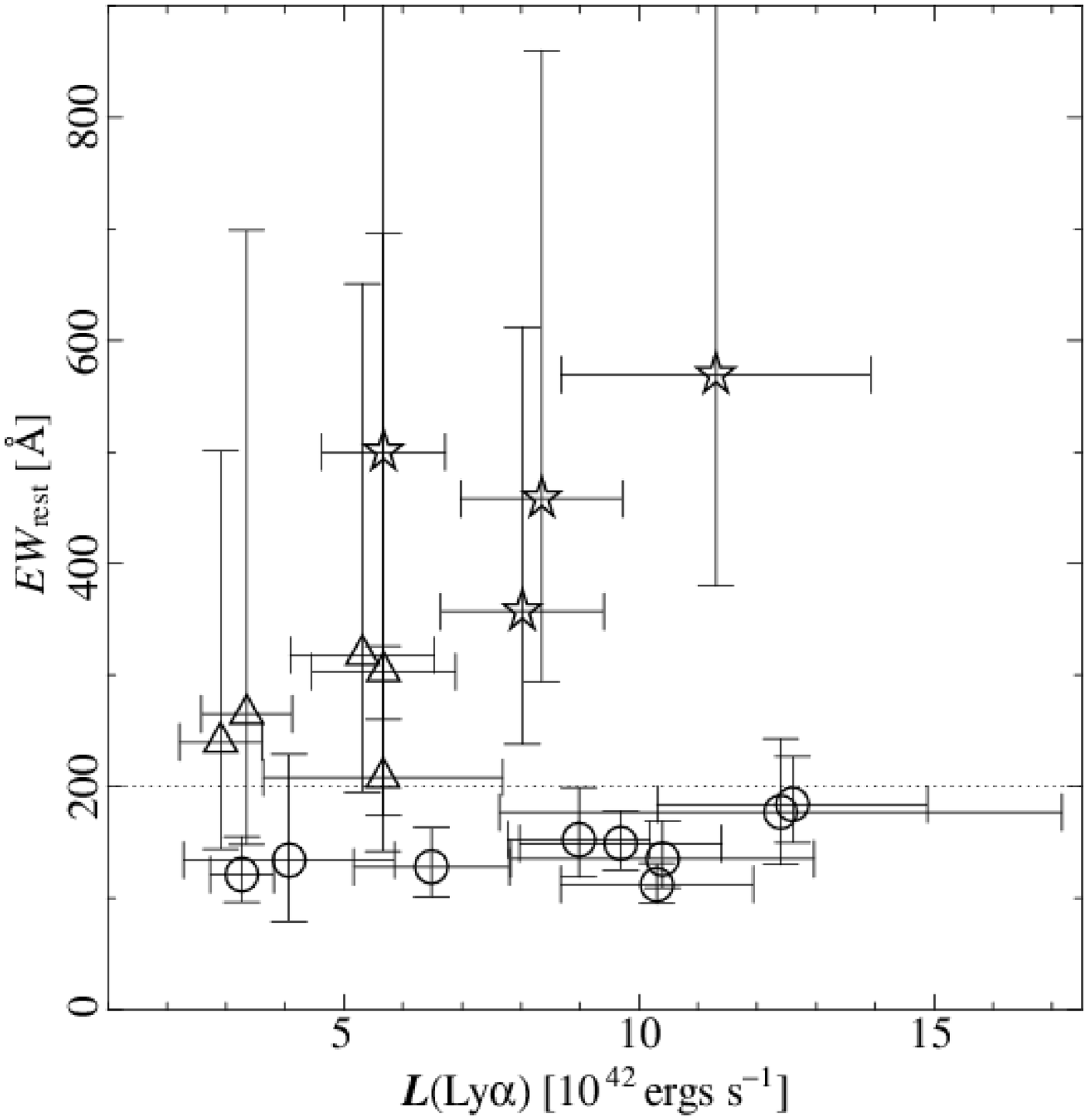}
\caption{Equivalent width of the \lya\ line as a function of the 
\lya\ line luminosity. The horizontal dotted line corresponds to 200\AA. 
Objects marked with open stars have EWs whose lower limits exceed 200\AA. 
Open triangles show objects with EWs larger than 200\AA, 
but with lower limits below 200\AA. The remaining objects, 
which have EWs less than 200\AA, are marked with open circles. 
}
\label{fig:lew}
\end{figure}

\begin{figure}
\plotone{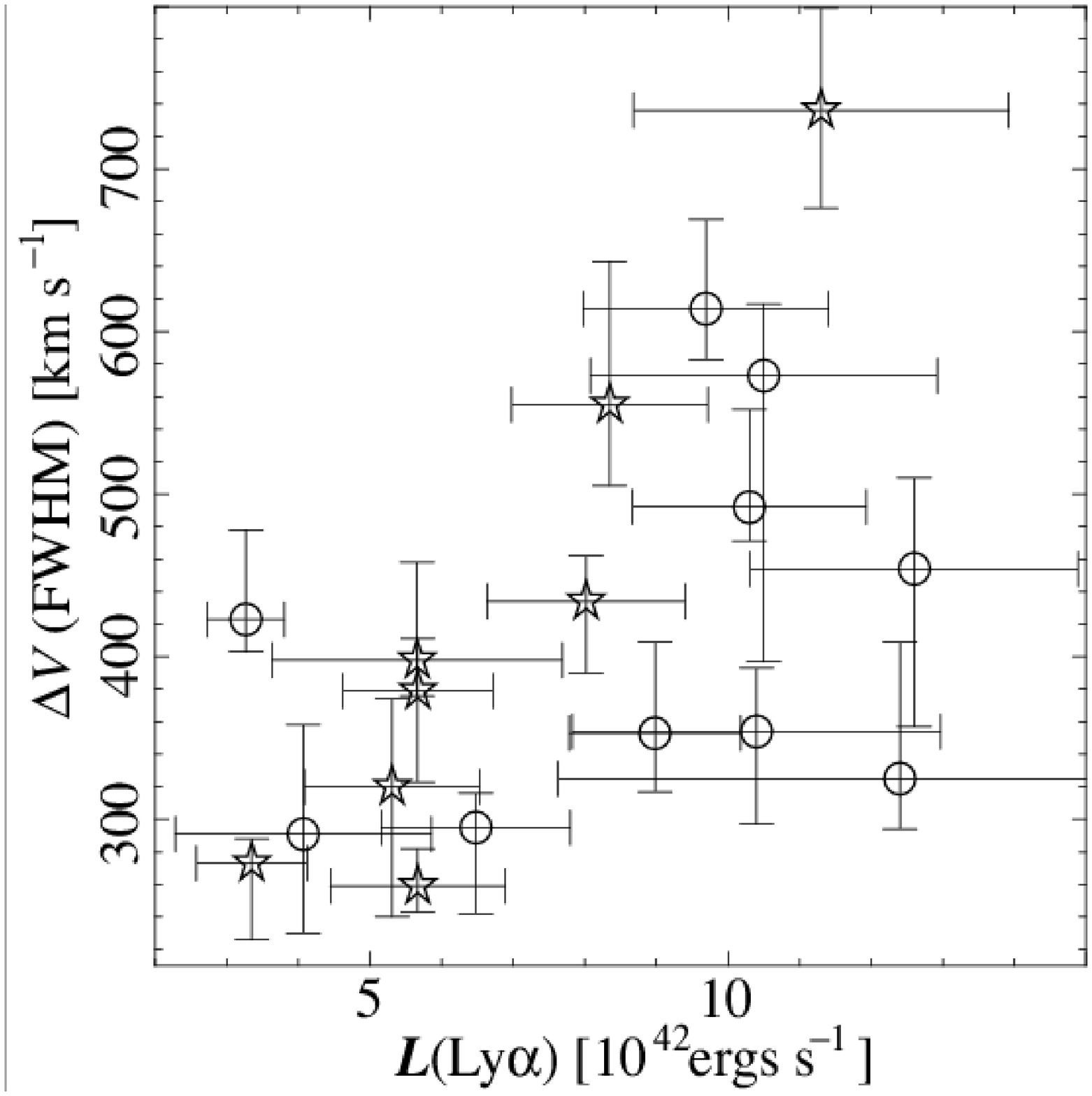}
\caption{Velocity width (FWHM) as a function of the \lya\ line 
luminosity. Objects with large EWs exceeding 200\AA\ are plotted 
with the open stars. 
}
\label{fig:ldv}
\end{figure}

\end{document}